\documentclass[10pt,letterpaper]{article}
\usepackage[top=0.85in,left=2.75in,footskip=0.75in]{geometry}

\usepackage{physics}

\usepackage{amsmath,amssymb}

\usepackage{changepage}

\usepackage[utf8x]{inputenc}

\usepackage{textcomp,marvosym}

\usepackage{cite}

\usepackage{nameref,hyperref}

\usepackage[right]{lineno}

\usepackage{microtype}
\DisableLigatures[f]{encoding = *, family = * }

\usepackage[table]{xcolor}

\usepackage{array}

\newcolumntype{+}{!{\vrule width 2pt}}

\newlength\savedwidth

\newcommand\thickhline{\noalign{\global\savedwidth\arrayrulewidth\global\arrayrulewidth 2pt}%
  \hline
  \noalign{\global\arrayrulewidth\savedwidth}}

\raggedright
\usepackage[aboveskip=1pt,labelfont=bf,labelsep=period,justification=raggedright,singlelinecheck=off]{caption}

\makeatletter
\renewcommand{\@biblabel}[1]{\quad#1.}
\makeatother

\usepackage{lastpage,fancyhdr,graphicx}
\usepackage{epstopdf}
\fancyheadoffset[L]{2.25in}
\fancyfootoffset[L]{2.25in}
\usepackage{dsfont}
\usepackage{chemfig,chemmacros}
\chemsetup{modules={reactions}}

\begin{document}
  \vspace*{0.2in}
  
  \begin{flushleft}
    {\Large
      \textbf\newline{Quantum Isomer Search} 
    }
    \newline
    \\
    Jason P. Terry\textsuperscript{1,2,3\dag},
    Prosper D. Akrobotu\textsuperscript{4,5\dag},
    Christian F. A. Negre\textsuperscript{5* },
    Susan M. Mniszewski\textsuperscript{6*}
    \\
    \bigskip
    \textbf{1} Department of Physics and Astronomy, University of Georgia, Athens, Georgia, United States of America
    \\
    \textbf{2} Data Science Initiative, Brown University, Providence, Rhode Island, United States of America
    \\
    \textbf{3} Center for Nonlinear Studies (T-CNLS), Theoretical Division, Los Alamos National Laboratory, Los Alamos, New Mexico, United States of America
    \\
    \textbf{4} Department of Mathematical Sciences, The University of Texas at Dallas, Richardson, Texas, United States of America
    \\
    \textbf{5} Physics and Chemistry of Materials (T-1), Theoretical Division, Los Alamos National Laboratory, Los Alamos, New Mexico, United States of America
    \\
    \textbf{6} Information Sciences (CCS-3), Computer, Computational and Statistical Sciences Division, Los Alamos National Laboratory, Los Alamos, New Mexico, United States of America
    \bigskip
    
    %
    %
    \dag These authors contributed equally to this work.
    
    
    
    
    *cnegre@lanl.gov,*smm@lanl.gov
    
  \end{flushleft}
  
  \section*{Abstract}
  
  Isomer search or molecule enumeration refers to the problem of finding all the isomers for a given molecule. Many classical search methods have been developed in order to tackle this problem. However, the availability of quantum computing architectures has given us the opportunity to address this problem with new (quantum) techniques. This paper describes a quantum isomer search procedure for determining all the structural isomers of alkanes. We first formulate the structural isomer search problem as a quadratic unconstrained binary optimization (QUBO) problem. The QUBO formulation is for general use on either annealing or gate-based quantum computers. We use the D-Wave quantum annealer to enumerate all structural isomers of all alkanes with fewer carbon atoms ($n<10$) than Decane ($C_{10}H_{22}$). The number of isomer solutions increases with the number of carbon atoms. We find that the sampling time needed to identify all solutions scales linearly with the number of carbon atoms in the alkane. We probe the problem further by employing reverse annealing as well as a perturbed QUBO Hamiltonian and find that the combination of these two methods significantly reduces the number of samples required to find all isomers. 
  
  
  \section*{Introduction}
  
  As quantum computers with more qubits and increased accuracy become available, interest in solving useful problems in the near-term has increased. The Ising problem is among this group. This problem is well-known to be NP-complete and is therefore efficiently mappable to all other NP-complete problems, such as the graph-coloring problem~\cite{ising_np}. Since both gate-based and annealing quantum computers can solve the Ising problem~\cite{gate_ising, anneal_ising}, it has been of particular interest to the quantum computing community for both near-term and long-term applications. It has already been successfully used to solve several problems such as graph partitioning and community detection~\cite{graph_partition, community_detection}.
  
  \par
  Given the structural formula of a molecule, one can always construct a graph defining the connectivity of the atoms. Thus the corresponding graph of a molecule, called a molecular graph, is defined as a labeled graph with a vertex set consisting of the atoms of the molecule and edges representing the chemical bonds existing between the atoms~\cite{Minkin1999}. For the case of hydrocarbons, we refer to the molecule as saturated if the bond between atoms is single and unsaturated if the bond is a double bond. In this vain, a saturated hydrocarbon is defined as a simple molecular graph whose vertices represent hydrogen and carbon atoms, and edges represent single bonds between the atoms~\cite{Minkin1999}. These simple molecular graphs could be either cyclic or acyclic. The acyclic saturated hydrocarbons are known as alkanes with molecular formula $C_{n}H_{2n+2}$ while the cyclic saturated hydrocarbons are referred to as the cycloalkanes with molecular formula $C_{n}H_{2n}$. A general formula for the saturated hydrocarbons is given as $C_{n}H_{2n-2(k-1)}$, where $k$ is the number of independent loops. A given molecular formula could correspond to different molecules with distinct structural arrangements. The molecules with identical formulas but distinct structures are called isomers. They are classified as structural isomers if their bonding patterns and atomic organization is distinct, or as stereoisomers if the bonding patterns are fixed while the spatial arrangement is distinct~\cite{Moss1996}. The goal of this work is to enumerate the structural isomers of any given molecular formula for alkanes by encoding this problem into a quantum computing framework.
  
  \par Isomer search, or molecule enumeration, is the process of searching for all isomers
  of a given molecule. The search space could be structural (2D) or spatial (3D), but for our purposes the focus will be on structural isomer search. The enumeration of structural isomers is of interest to numerous fields. Examples include pharmaceutical applications as proper identification of isomers ``would facilitate resource reduction, including animal usage, and may benefit other areas of pharmaceutical structural characterization including impurity profiling and degradation chemistry"~\cite{pharma_isomer}. Moreover, the oil and gas industry relies on knowledge of the structures of isomers of hydrocarbons and alkanes for processes related to refinement~\cite{hydroisomerization, cracking_isomers}
  
  \par
  Over the years, many search procedures have been developed for enumerating molecules. Most of these procedures incorporate various techniques such as the labeled enumeration (a procedure for enumerating all $2^{\frac{(n(n-1))}{2}}$ labeled graphs of $n$ nodes), orderly generation method (a procedure that dwells on a so-called canonical representation of graphs such that the canonization process induces an ordering on the edges of the graph), random sampling (techniques that generate structures from randomly selected branches of the construction trees of deterministic structural generation algorithms), Monte-Carlo and simulated annealing (a procedure that focuses on minimizing random displacements on atoms by performing a bond order switch), and genetic algorithms (a procedure for which mutations are carried out using bond perturbations, crossover operations executed using a generated $n$-tuple code, and selection operators defined by root-mean-square deviation between experimental chemical shifts, and predicted chemical shifts from neural network technology)~\cite{EnumeratingMolecules}.
  
  \par
  The work of Nobel Laureate J. Lederberg~\cite{Lederberg_topology} on the topology of molecules in 1969 is considered to be the genesis of algorithmic and computational approaches to the field of molecular structure enumeration. A particularly large step was when the algorithm was finally incorporated into the DENDRAL (Dendritic Algorithm) code~\cite{Dendral_project} for enumerating isomers of acyclic compounds containing carbon (C), hydrogen (H), oxygen (O), and nitrogen (N) atoms~\cite{Acyclic_structures_CHON}. Since then, the field has seen the development of several distinct types of algorithms (exhaustive, automated and stochastic) and codes for structural elucidation using different classical approaches with diverse input criteria~\cite{EnumeratingMolecules}. In order to generate the isomers, most of these codes, including SKELGEN, CAMGEC, AEGIS, ISOGEN, GI, and MOLGEN~\cite{Skel_gen, Camgec,Aegis, Isogen, GI, Molgen}
  require only the molecular formula as input. Others,
  such as DENDRAL, GalvaStructures, CONGEN, GENOA , ASSEMBLE 2.0, CHEMICS, EPIOS, GEN, StructEluc, COCON, SpecSolv, ESESOC, SIGNATURE, SENECA, COCOA, SESAMI, CISOC-SES, and X-PERT
  take as input any combination of molecular formula, spectral information and spectroscopic data, fragments, molecular weight, molar mass, and constraints such as long-range distance constraints~\cite{Dendral_project, Le_Bret, Congen, Genoa, Assemble, Chemics, Epios, Gen, Structeluc, Cocon, Specsolv, Esesoc1, Esesoc2, Signature1, Signature2, Seneca, Cocoa, Sesami, Cisocses, Xpert}.
  
  \par
  Here we introduce, formulate, and apply quantum isomer searching. Based on the graph-coloring problem, and formulated as a QUBO/Ising model, our approach is able to identify structural isomers of a given molecule in a way that can be implemented on both quantum annealers and gate-based quantum computers. This particular model is able to search for structural isomers of alkanes. 
  
  \par
  In order to validate our formulation, the search is implemented on the D-Wave 2000Q machine, a state-of-the-art quantum annealer with 2048 superconducting qubits arranged in a sparse chimera graph~\cite{DWave2000Q}. It is a quantum computing device that works using quantum annealing, a method that makes use of quantum tunneling and quantum entanglement in order to solve combinatorial optimization problems through minimizing the Ising objective function:
  \begin{align}
    \label{eq:ising}
    f(\boldsymbol{\sigma}) = \sum_{i}h_{i}\sigma_{i} + \sum_{i<j}J_{i,j}\sigma_{i}\sigma_{j}
  \end{align}
  for which $\sigma_{i}\in\{-1,1\}$ are magnetic spin variables subject to local fields $h_{i}$ and nearest neighbor interactions with coupling strength $J_{i,j}$. Any problem to be solved on a D-Wave system is modeled as a search for the minimal energy of the Ising Hamiltonian. When the variables $\sigma_{i}$ in Eq ~\ref{eq:ising} are restricted to take values from the set  $\{0,1\}$, then the minimization problem is said to be a quadratic unconstrained binary optimization (QUBO).  A typical QUBO model can be transformed into an Ising model with the transformation $\boldsymbol{\sigma} = 2\boldsymbol{x} - \mathds{1}_{n}$, where entries of $\boldsymbol{x}$ represent the $n$ QUBO variables and $\mathds{1}_{n}$ is a vector of ones. In the following sections, we formulate the quantum isomer search problem as a QUBO, then describe and present its implementation.
  
  \par
  In our formulation, an alkane with $n$ carbons requires $4(n-2)$ logical qubits that are fully connected. However, D-Wave 2000Q's chimera graph is sparse, and therefore may require a logical qubit to be represented by a chain of physical qubits~\cite{chimera}. 
  This architecture can limit the complexity of possible problems and creates the difficult task of mapping the necessary connections of the logical qubits onto the possible connections of the physical qubits in a process known as ``minor embedding''~\cite{chimera}. This leads to the D-Wave 2000Q being capable of representing up to 64 fully connected logical qubits or variables, meaning that our method can find isomers of alkanes with up to $n\leq18$ carbon atoms (Octadecane). 
  
  \par
  An important aspect of this problem 
  is that there are multiple correct answers for a given QUBO, i.e. more than one answer satisfies all given constraints. In the case of D-Wave 2000Q, this means that all of these answers have the global minimum energy. In this way, the ground state is degenerate, and to fully answer a given problem, all answers with that energy must be found. This degeneracy is an essential, and necessary part of the problem since the purpose is to find all valid solutions. Quantum annealers are ideal for sampling degenerate solutions because of their ability to introduce some randomness in their exploration of the search space. However, this creates complications because it requires the search space to be explored to an exhaustive degree, which quickly becomes more difficult as the problem size increases. This is a well-known issue with annealing devices, and previous results have found that it can be difficult to sample all degenerate solutions in a fair way~\cite{degeneracy}. To this end, we apply techniques in an attempt to encourage the search space to be more fully explored than it is with a typical anneal.
  
  \section*{Methods}
  
  \subsection*{QUBO Formulation}
  
  Employing acyclic molecular tree graphs to represent alkanes, as well as considering the specific properties of these graphs, we formulate isomer search as a quadratic unconstrained binary optimization (QUBO) problem that can be solved via quantum annealing or gate-based quantum computers. We start off this section by constructing the QUBO objective function for searching for the isomers based on their degree sequences.
  \\~\\
  Given a molecular formula for an alkane, $C_{n}H_{2n+2}$, we consider the carbon-carbon connectivity and set up a degree sequence $(x_{1},x_{2},\ldots,x_{n})$ of the corresponding acyclic molecular tree graph.  For our purposes, the hydrogen atoms are irrelevant because they can be inferred from the arrangement of the carbons and can therefore be dropped from the graph. Using the constraints on degree sequences of tree graphs we construct the QUBO objective function of the form
  \begin{align*}
    \boldsymbol{x}^{T}Q\boldsymbol{x} = \sum_{i=1}^{n}x_{i}Q_{ii} + \sum_{i\neq j}x_{i}x_{j}Q_{ij}
  \end{align*} 
  where each element $x_{i}$ of the vector $\boldsymbol{x}$  belongs to the set $\{0,1\}$.  
  Recall that trees are such that at least two nodes are of degree $1$. Without loss of generality, these nodes can be reordered such that they are located at the first and last positions of the corresponding degree sequence. This enables us to set a constraint as $x_{1}=x_{n}=1$. These nodes correspond to carbons that are in methyl groups ($CH_{3}$). Furthermore, the carbon-carbon bond in alkane is such that each carbon atom is bonded to at most 4 other carbon atoms. This gives rise to another constraint $1\leq x_{j}\leq 4 \hspace{10pt} \text{for } j=2,\ldots, n-1$.  By the properties of trees, we establish that the sum of the degree sequences must be $2(n-1)$, i.e. $x_{1}+x_{2}+\ldots + x_{n} = 2(n-1)$. Putting all these together gives:
  
  \begin{align}
    \label{degConstraint1}
    &  x_{1}=x_{n}=1\\
    \label{degConstraint2}
    &1\leq x_{j}\leq 4 \hspace{10pt} \text{for } j=2,\ldots, n-1\\
    \label{degConstraint3}
    &\sum_{j=2}^{n-1}x_{j} = 2(n-2)    
  \end{align}
  
  To convert the $x_{i}$ to binary, we define decision variable $y_{ij}$ based on the graph-coloring idea for which a node $i$ is assigned a color $j$, by considering degrees as colors, that is $j=1,\ldots,4$ since the maximum degree is 4. In other words, the number of carbon bonds for an individual atom is one hot encoded in a bit string of length 4.
  \begin{align*}
    y_{ij} = \begin{cases} 1 & \text{if node }  i \text{ is of degree } j (x_{i} = j)\\ 0& \text{otherwise }\end{cases}
  \end{align*}
  For these variables, we can establish the constraints : 
  \begin{align}
    \label{eq:constraint1}
    \sum_{j=1}^{4}y_{ij} &= 1 \hspace{20pt} i=1,\ldots,n\\
    y_{11} = y_{n1} &= 1\nonumber\\
    \label{eq:constraint2}
    \sum_{i=1}^{n}\sum_{j=1}^{4}jy_{ij} &= 2(n-1)     
  \end{align}

  \par
\begin{figure}[h!]
    \centering
    \includegraphics[width=0.75\linewidth]{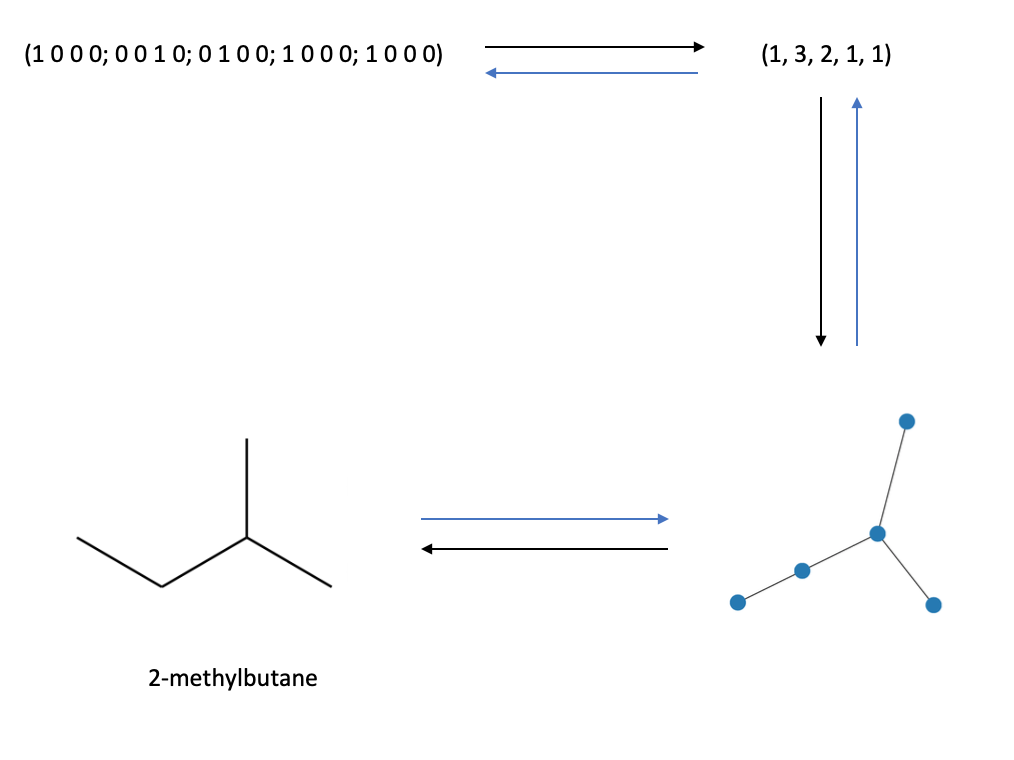}
    \caption{{\bf Encoding of 2-methylbutane.}
      Representation of 2-methylbutane (an isomer of pentane $C_{5}H_{12}$) as a one hot encoded bit string, degree sequence, graph, and molecular graph (clockwise from the top left).}
    \label{encoding}
  \end{figure}

  Fig.~\ref{encoding} shows, using 2-methylbutane (an isomer of pentane $C_{5}H_{12}$) as an example, how a given alkane can be represented as a molecular graph, a tree graph, a degree sequence, and a one hot encoded bit string. It is important to note that the order of the returned degree sequence is physical. For alkanes with $n\geq6$ carbons, there are multiple valid permutations of the same degree sequence that lead to different isomers. Alternatively, there are often different permutations that create equivalent isomers. Therefore, this encoding requires post-processing steps to be taken in order to make these distinctions.
  
  \par
  In this new formulation, there are $4n$ variables, $8$ of which are already predetermined as a result of the constraint $y_{11}=y_{n1}=1$. Thus, we can restrict the problem to only $M = 4(n-2)$ variables. This, not only reduces computational complexity but also enables us to explore larger alkanes due to the restrictions on the number of variables or qubits on current machines. For the D-Wave 2000Q machine, this means the ability to explore alkanes with up to 18 carbon atoms instead of 16 carbon atoms.
  
  For simplicity, let us re-number the indices as
  \begin{align*} \boldsymbol{y}&=(y_{11},y_{12},y_{13},y_{14},y_{21},y_{22},\ldots,y_{(n-2)1},y_{(n-2)2},y_{(n-2)3},y_{(n-2)4})\\ &= (y_{1},y_{2},\ldots,y_{M})\end{align*}
  
  We introduce positive penalty constants $P_{i}$ and proceed with the construction of the QUBO by following the methods in \cite{DBLP:journals/corr/abs-1811-11538}. We penalize the $n$ constraints in Eq~\ref{eq:constraint1}  with penalty constant $P_{1}$ and the constraint in Eq~\ref{eq:constraint2} with penalty constant $P_{2}$ as follows:
  
  \begin{align}
    P_{1}\left(\sum_{i=1}^{4}y_{i}-1\right)^{2} &=P_{1}\left(-\sum_{i=1}^{4}y_{i} + 2\sum_{1\leq i<j\leq 4}y_{i}y_{j}\right) + P_{1}\nonumber\\
    P_{1}\left(\sum_{i=1}^{4}y_{4+i}-1\right)^{2} &=P_{1}\left(-\sum_{i=1}^{4}y_{4+i} + 2\sum_{1\leq i<j\leq 4}y_{4+i}y_{4+j}\right) + P_{1}\nonumber\\
    P_{1}\left(\sum_{i=1}^{4}y_{8+i}-1\right)^{2} &=P_{1}\left(-\sum_{i=1}^{4}y_{8+i} + 2\sum_{1\leq i<j\leq 4}y_{8+i}y_{8+j}\right) + P_{1}\nonumber\\
    \vdots \nonumber\\
    P_{1}\left(\sum_{i=1}^{4}y_{4(n-3)+i}-1\right)^{2} &=P_{1}\left(-\sum_{i=1}^{4}y_{4(n-3)+i} + 2\sum_{1\leq i<j\leq 4}y_{4(n-3)+i}y_{4(n-3)+j}\right) + P_{1}\nonumber\\
    \label{eqn:penaltyP2}
    \Longrightarrow P_{1}\sum_{j=0}^{n-3}\left(\sum_{i=1}^{4}y_{4j+i}-1\right)^{2} &= P_{1} \left[-\sum_{i=1}^{M}y_{i} +2\sum_{k=0}^{n-3}\sum_{1\leq i<j\leq 4}y_{4k+i}y_{4k+j} + (n-2)\right]\\
    \label{eqn:penaltyP4}
    P_{2}\left(\sum_{i=1}^{n-2}\sum_{j=1}^{4}jy_{ij} - 2(n-2)\right)^{2}
    &=P_{2}\left[\sum_{i=1}^{M}[(i-1)(\mod 4) +1]y_{i} - 2(n-2)\right]^{2}
  \end{align}
  
  Let $\mathds{1}$ be the $M\times M$ matrix of ones and $\boldsymbol{\mathds{1}_{M}}$ be the $M$ column vector of ones. Let $U = [u_{ij}]_{4\times 4}$ be a $4\times 4$ upper triangular matrix with entries defined by
  \begin{align*}
    u_{ij} = \begin{cases}1 & \text{if }i<j\\ 0 & \text{ otherwise}\end{cases}.
  \end{align*}
  That is, 
  \begin{align*}
    U = \begin{bmatrix}0 & 1 & 1 & 1\\0&0&1&1\\0&0&0&1\\0&0&0&0 \end{bmatrix}.
  \end{align*}
  We define a $M \times M$ (block) diagonal matrix $D_{U}$ with each diagonal block consisting of the matrix $U$ as 
  \begin{align*}
    D_{U} = diag (U,\ldots,U).
  \end{align*}
  
  Eq~\ref{eqn:penaltyP2} can then be rewritten as
  \begin{align}
    P_{1}\sum_{j=0}^{n-3}\left(\sum_{i=1}^{4}y_{4j+i}-1\right)^{2} &= P_{1} \left[-\sum_{i=1}^{M}y_{i} +2\sum_{k=0}^{n-3}\sum_{1\leq i<j\leq 4}y_{4k+i}y_{4k+j} + n\right]\nonumber\\
    \label{eqn:penaltyP2expanded}
    &=P_{1}\left[-\boldsymbol{\mathds{1}}_{M}^{T}\boldsymbol{y} + 2\boldsymbol{y}^{T}D_{U}\boldsymbol{y} + (n-2)\right].
  \end{align}
  
  To rewrite Eq~\ref{eqn:penaltyP4} in matrix form, we first define $\alpha_{i} = (i-1)\mod 4 + 1$. Then
  \begin{align}
    &P_{2}\left[\sum_{i=1}^{M}[(i-1)\mod 4 +1]y_{i} - 2(n-2)\right]^{2}\nonumber\\ &=P_{2}\left[ \sum_{i=1}^{M}\alpha_{i}y_{i} - 2(n-2)\right]^{2}\nonumber\\
    &=P_{2}\left[\left(\sum_{i=1}^{M}\alpha_{i}y_{i}\right)^{2} - 4(n-2)\sum_{i=1}^{M}\alpha_{i}y_{i} + 4(n-2)^{2}\right]\nonumber\\
    \label{eqn:penaltyP4epanded1}
    &=P_{2}\left[\sum_{i=1}^{M}\alpha_{i}^{2}y_{i}^{2}+ 2\sum_{1\leq i<j\leq M}\alpha_{i}\alpha_{j}y_{i}y_{j} - 4(n-2)\sum_{i=1}^{M}\alpha_{i}y_{i} + 4(n-2)^{2}\right].
  \end{align}
  
  Now, define the following $M\times M$ matrices and $M$ column vectors
  \begin{align}
    D_{\alpha} &= diag(\alpha_{1},\ldots,\alpha_{M})\\
    \boldsymbol{\alpha} &= (\alpha_{1},\ldots,\alpha_{M}) \\
    U_{\alpha} &= [u_{ij}]_{M\times M} \hspace{10pt}\text{with } u_{ij} = \begin{cases}\alpha_{j} & \text{if } i<j \\ 0 & \text{ otherwise}\end{cases}
  \end{align}
  Then
  \begin{align}
    &P_{2}\left[\sum_{i=1}^{M}\alpha_{i}^{2}y_{i}^{2}+ 2\sum_{1\leq i<j\leq M}\alpha_{i}\alpha_{j}y_{i}y_{j} - 4(n-2)\sum_{i=1}^{M}\alpha_{i}y_{i} + 4(n-2)^{2}\right]\nonumber\\
    \label{eqn:penaltyP4exapnded}
    &=P_{2}\left[\boldsymbol{y}^{T}D_{\alpha}^{2}\boldsymbol{y} + 2\boldsymbol{y}^{T}D_{\alpha}U_{\alpha}\boldsymbol{y} - 4(n-2)\boldsymbol{\alpha}^{T}\boldsymbol{y} + 4(n-2)^{2}\right] 
  \end{align}
  
  Adding all these equations, Eq~\ref{eqn:penaltyP2expanded}  + Eq~\ref{eqn:penaltyP4exapnded}, gives
  
  \begin{align}
    &P_{1}\left[-\boldsymbol{\mathds{1}}_{M}^{T}\boldsymbol{y} + 2\boldsymbol{y}^{T}D_{U}\boldsymbol{y} + (n-2)\right]
    + P_{2}\left[\boldsymbol{y}^{T}D_{\alpha}^{2}\boldsymbol{y} + 2\boldsymbol{y}^{T}D_{\alpha}U_{\alpha}y - 4(n-2)\boldsymbol{\alpha}^{T}\boldsymbol{y} + 4(n-2)^{2}\right]\nonumber\\
    &=\boldsymbol{y}^{T}\left[P_{2}D_{\alpha}^{2} +2(P_{2}D_{\alpha}U_{\alpha} + P_{1}D_{U})\right]\boldsymbol{y} 
    + \left[-4(n-2)P_{2}\boldsymbol{\alpha} - P_{1}\boldsymbol{\mathds{1}}_{M}) \right]^{T}\boldsymbol{y}\nonumber\\
    &+ \left[4(n-2)^{2}P_{2} + (n-2)P_{1} \right]\\
    \label{eqn:qubo_objectivefxn}
    &= \boldsymbol{y}^{T}A\boldsymbol{y} + \boldsymbol{b}^{T}\boldsymbol{y} + c,
  \end{align}
  where 
  \begin{align}
    \label{eqn:matrixA}
    A &= P_{2}D_{\alpha}^{2} + 2(P_{2}D_{\alpha}U_{\alpha} + P_{1}D_{U}),\\
    \label{eqn:vectorb_simplified}
    \boldsymbol{b} &= -\left[4(n-2)P_{2}\boldsymbol{\alpha} + P_{1}\boldsymbol{\mathds{1}}_{M} \right],\\
    \label{eqn:number_c}
    c &= 4(n-2)^{2}P_{2}  + (n-2)P_{1}.
  \end{align}
  
  Eq~\ref{eqn:qubo_objectivefxn} with $A,\boldsymbol{b},c$ defined by Eq~\ref{eqn:matrixA} --  Eq~\ref{eqn:number_c} respectively, gives us the objective function for our problem and hence we need to solve the minimization problem
  \begin{align}
    \label{eqn:minproblem}
    \min_{\boldsymbol{y}\in \{0,1\}^{M}}\boldsymbol{y}^{T}A\boldsymbol{y} + \boldsymbol{b}^{T}\boldsymbol{y} + c.
  \end{align}
  
  Let us define a diagonal matrix $D_{\boldsymbol{b}}$ out of the vector $\boldsymbol{b}$ above, that is 
  \begin{align*}
    D_{\boldsymbol{b}} = diag(b_{1},b_{2},\ldots,b_{M}) 
  \end{align*}
  where the $b_{i}$ are the vector components of $\boldsymbol{b}$ in Eq~\ref{eqn:vectorb_simplified}.
  Then we solve the QUBO problem
  \begin{align}
    \label{eqn:QUBO_min}
    \min_{\boldsymbol{y}\in \{0,1\}^{M}}\boldsymbol{y}^{T}A\boldsymbol{y} + \boldsymbol{b}^{T}\boldsymbol{y}
    &=\min_{\boldsymbol{y}\in \{0,1\}^{M}}\boldsymbol{y}^{T}A \boldsymbol{y} + \boldsymbol{y}^{T}D_{\boldsymbol{b}}\boldsymbol{y}\hspace{10pt} \text{since }y_{i}^{2}=y_{i}\in\{0,1\}\nonumber\\
    &= \min_{\boldsymbol{y}\in \{0,1\}^{M}}\boldsymbol{y}^{T}Q\boldsymbol{y} \hspace{20pt} \text{ where } Q = A+D_{\boldsymbol{b}}
  \end{align}
  
  \par
  In terms of the individual variables, the full QUBO objective function can be expressed as:
  
  \begin{align}
    \label{eqn:objective_function}
    f(\boldsymbol{y}) =\min_{\boldsymbol{y}\in \{0,1\}^{M}} P_{1}\sum_{i = 2}^{n - 1}\left(\sum_{j = 1}^{4}y_{i, j} -1 \right)^{2} + P_{2}\left(\sum_{i = 2}^{n - 1} \sum_{j = 1}^{4} jy_{i, j}- 4(n-2)\right)^{2}.
  \end{align}
  
  \par
  However, D-Wave 2000Q only accepts $h_{i}$ values between -2 and 2 and $J_{i,j}$ values between -1 and 1, so the resulting coefficients must be scaled to fit within these ranges. 
  This can help increase the energy gap between the ground state and the first excited state, making it less likely that excited states will be sampled.
  While the \texttt{D-Wave Ocean} software~\cite{ocean} can do this on its own, it was done manually since this allowed some experimentation with the range used as it has been observed that it may sometimes be useful to restrict $J_{i,j}$ values to be greater than -0.8~\cite{hj_range}. This was done, although no significant benefit was observed. 
  
  \par
  
\begin{figure}[!h]
    \centering
    \includegraphics[width=0.4\linewidth]{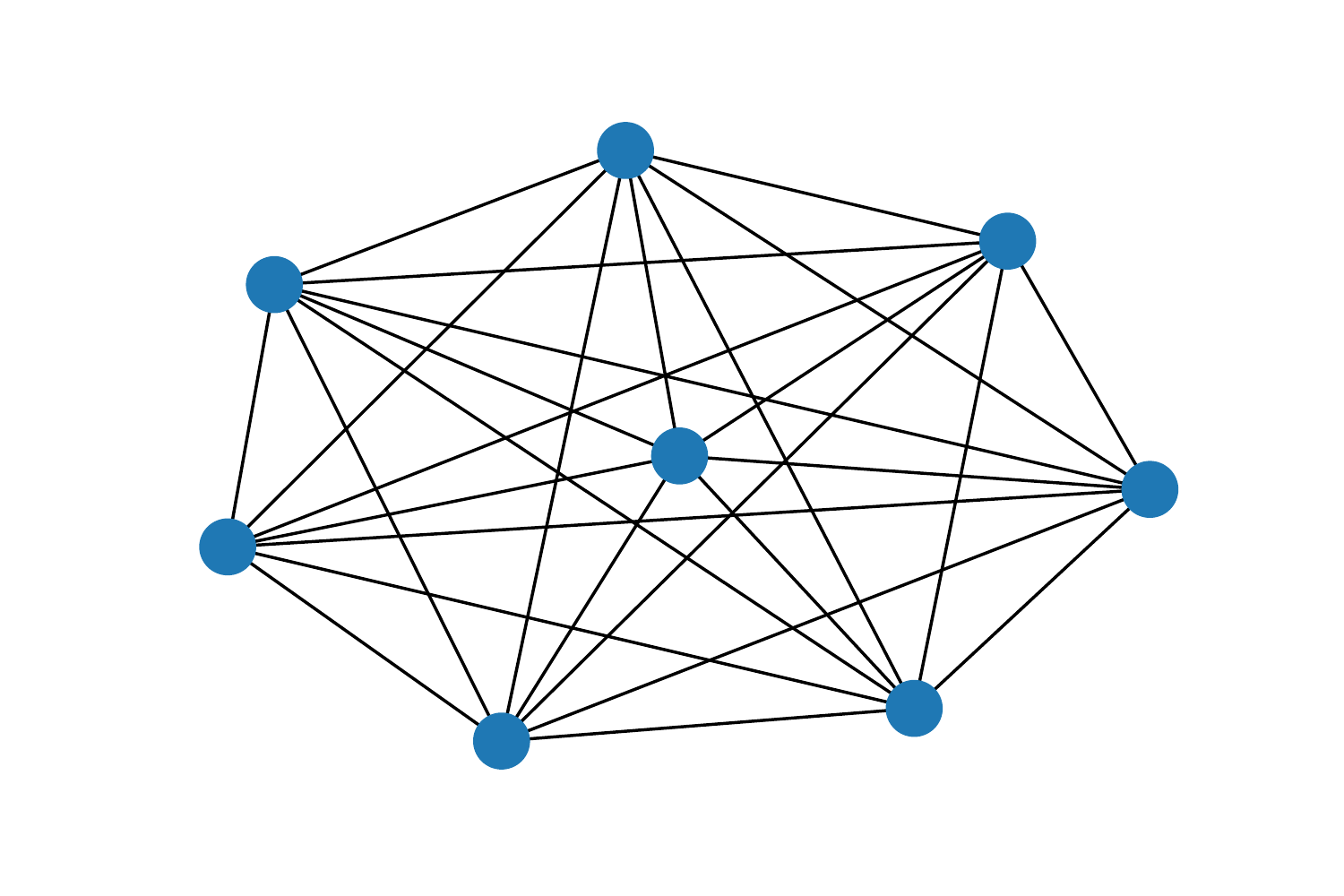}
    \includegraphics[width=0.4\linewidth]{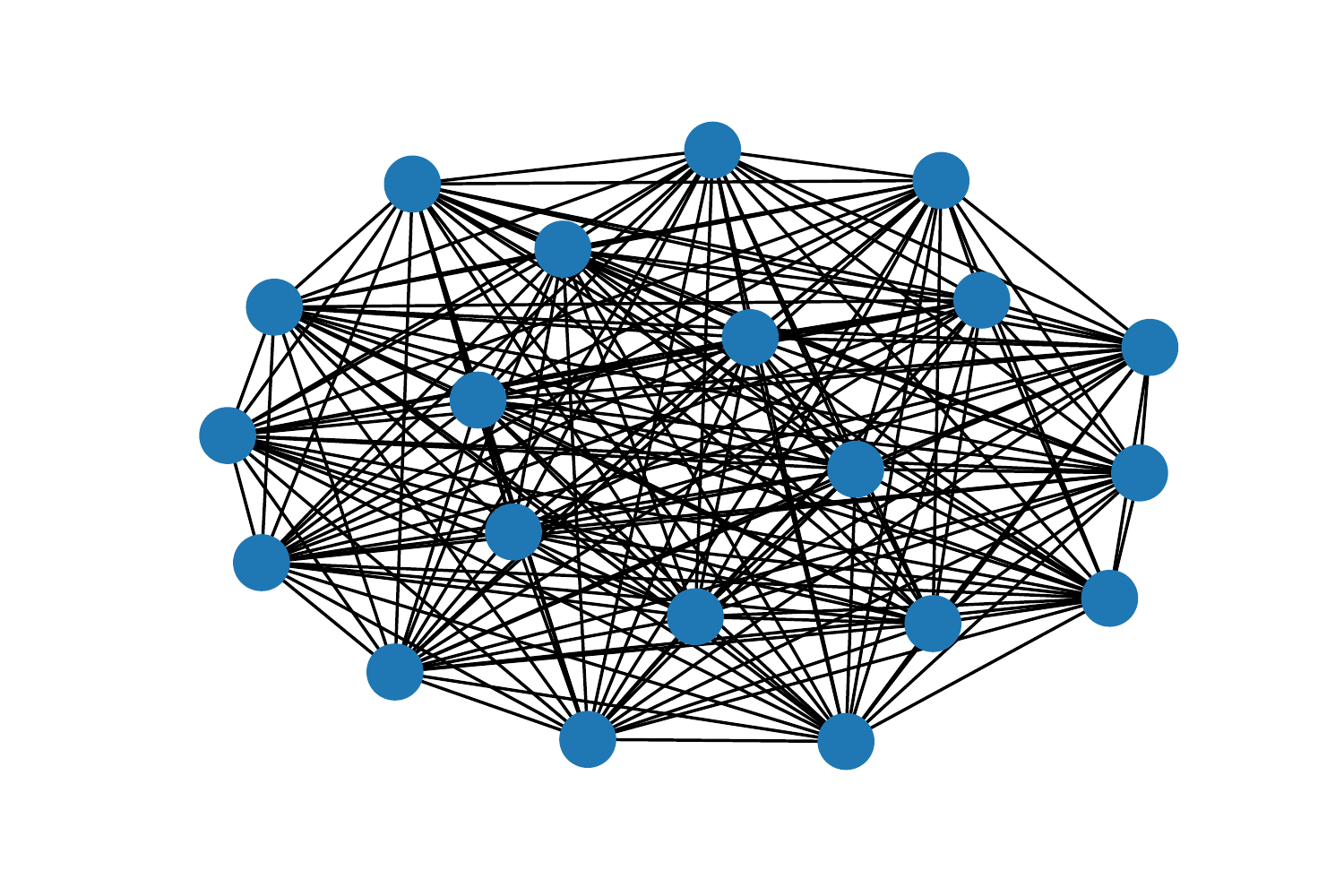}
    \caption{{\bf Graphs of QUBOS.}
      A: Butane ($C_{4}H_{10}$), B: Heptane ($C_{7}H_{16}$)}
    \label{qubos}
  \end{figure}

  It is important to note that, as a result of the constraint relating to the sum of the degrees of the atoms, the resulting QUBO is fully connected. Each atom must take into account the number of carbon bonds of all other atoms in order to determine if its coloring violates this constraint. This can be visualized using the graph representations of the QUBOs (which is what would be embedded into the D-Wave chimera graph) for Butane ($n=4$) and Heptane ($n=7$) given in Fig.~\ref{qubos}
  
  \subsection*{Implementation}
  
  The QUBO for a given alkane is embedded into the D-Wave 2000Q\_5 chimera graph. This is the newly released lower-noise machine that is available via D-Wave's Leap\textsuperscript{TM} quantum cloud service~\cite{2000q}. It was also implemented on the D-Wave 2000Q\_LANL machine at Los Alamos National Laboratory~\cite{dwave_lanl}, but no significant difference in performance were noted. Once the QUBO is embedded using \texttt{D-Wave Ocean}~\cite{ocean}, the annealer attempts to find the lowest energy solutions, i.e. the bit strings that violate the fewest constraints. This was done using the standard 20$\mu s$ anneal time. It is important to remember that the lowest energy solutions found by annealing are the valid isomers. Note that this is different from finding the most chemically stable isomer. 
  
  \par
  The sampled results are then filtered such that only the lowest energy solutions (the isomers) are returned. These one hot encoded results are decoded into the degree sequences and graphs in the method described previously, checked and filtered for redundancy, and returned. As $n$ increases, the relative number of possible results ($2^{4(n - 2)}$) grows more quickly than the number of isomers. Furthermore, it is known that larger problems on imperfect quantum annealers have lower probabilities of sampling a ground state solution~\cite{error}. Therefore, it becomes necessary to increase the number of samples taken from the embedded QUBO.
  
  \par
  To address this problem, an increasingly perturbed QUBO was also used. In this formulation, after every 10,000 samples, the outer product of the ground state result with the most counts, $\ket{\psi}$, with itself was added to the QUBO in an attempt to impose a penalty on returning that result in the next iteration. This new QUBO is represented as 
  
  \begin{align}
    \label{eqn:Q_prime}
    Q' = Q + \lambda \ket{\psi}\bra{\psi}.
  \end{align}
  
  The idea behind using this perturbed QUBO is that, by penalizing previously returned results, subsequent sampling runs would be encouraged to explore different parts of the search space that may have valid solutions that had not been visited. Because the search space becomes extremely large as $n$ increases, it is possible that this may help facilitate the identification of all isomers of larger molecules. To the best of our knowledge this is the first time that such a technique has been used to boost the solution space exploration in quantum annealers.
  
  \par
  Finally, reverse annealing was added. Rather than ending in a classical state after slowly turning down the strength of the transverse field, this method does the opposite by taking a fully classical state as input, which is then stimulated with an increasingly high transverse field until it reaches the pause location, $s^{*}$~\cite{reverse_anneal}.  At this location, $\mathcal{H}(t) = \mathcal{H}(s^{*})$, where $\mathcal{H}(0)$ and $\mathcal{H}(1)$ are the starting and ending Hamiltonians of a forward anneal, respectively. Following this step, the system pauses for some pre-determined time and progresses as a typical forward anneal as the transverse field is gradually weakened, eventually ending once again in a classical state ~\cite{reverse_anneal}. The classical input states were the results given by a typical forward anneal, the pause location was chosen to be $s^{*}=0.5$, and the system was paused for $h=85 \mu s$. One of the ideas behind reverse annealing is that it allows the search space surrounding candidate solutions given by a forward anneal to be further explored~\cite{reverse_anneal}. If the forward anneal returns a local minimum then reverse annealing may stimulate that solution to an extent that the system settles into a nearby global minimum~\cite{reverse_anneal}. Such a result could make it more likely that a given run returns a result with the minimum energy, which may help with the successful enumeration of all isomers and even decrease the number of samples necessary to find them all.
  
  \par
  Since gate-based quantum computers can also solve the Ising Problem~\cite{gate_ising}, in addition to using D-Wave 2000Q, we explored the possibility of using IBM Q's \texttt{Qiskit} software on the available QASM simulator~\cite{qiskit} . A QUBO can be expanded into the Pauli basis, and when this is done, it can then be solved using methods such as variational quantum eigensolver (VQE)~\cite{vqe} or quantum approximate optimization algorithm (QAOA)~\cite{qaoa}. Because of the requirement of $4(n-2)$ qubits, IBM Q's Tokyo, which has 20 qubits and is currently their largest available device, can only handle alkanes with fewer carbon atoms than Octane ($C_{8}H_{18}$)~\cite{ibm_hardware}. However, Google Bristlecone's 72 qubits will be able to encode 18 carbon atoms, allowing Icosane's ($n=20$) 366,319 structural isomers to be searched for~\cite{bristlecone}. 
  
  \section*{Results}
\begin{figure}[!h]
    \centering
    \includegraphics[width=0.48\linewidth]{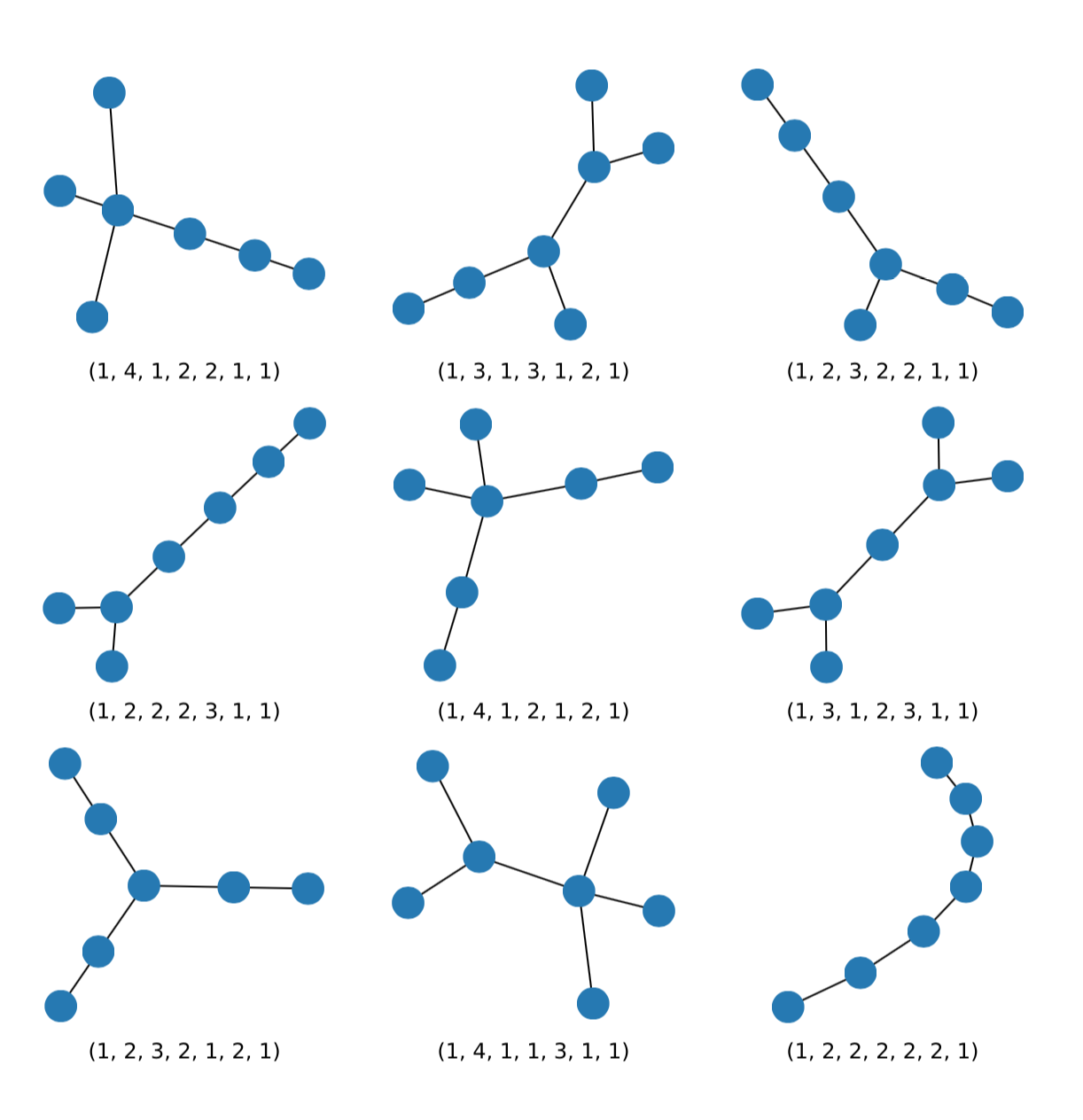}
    \includegraphics[width=0.512\linewidth]{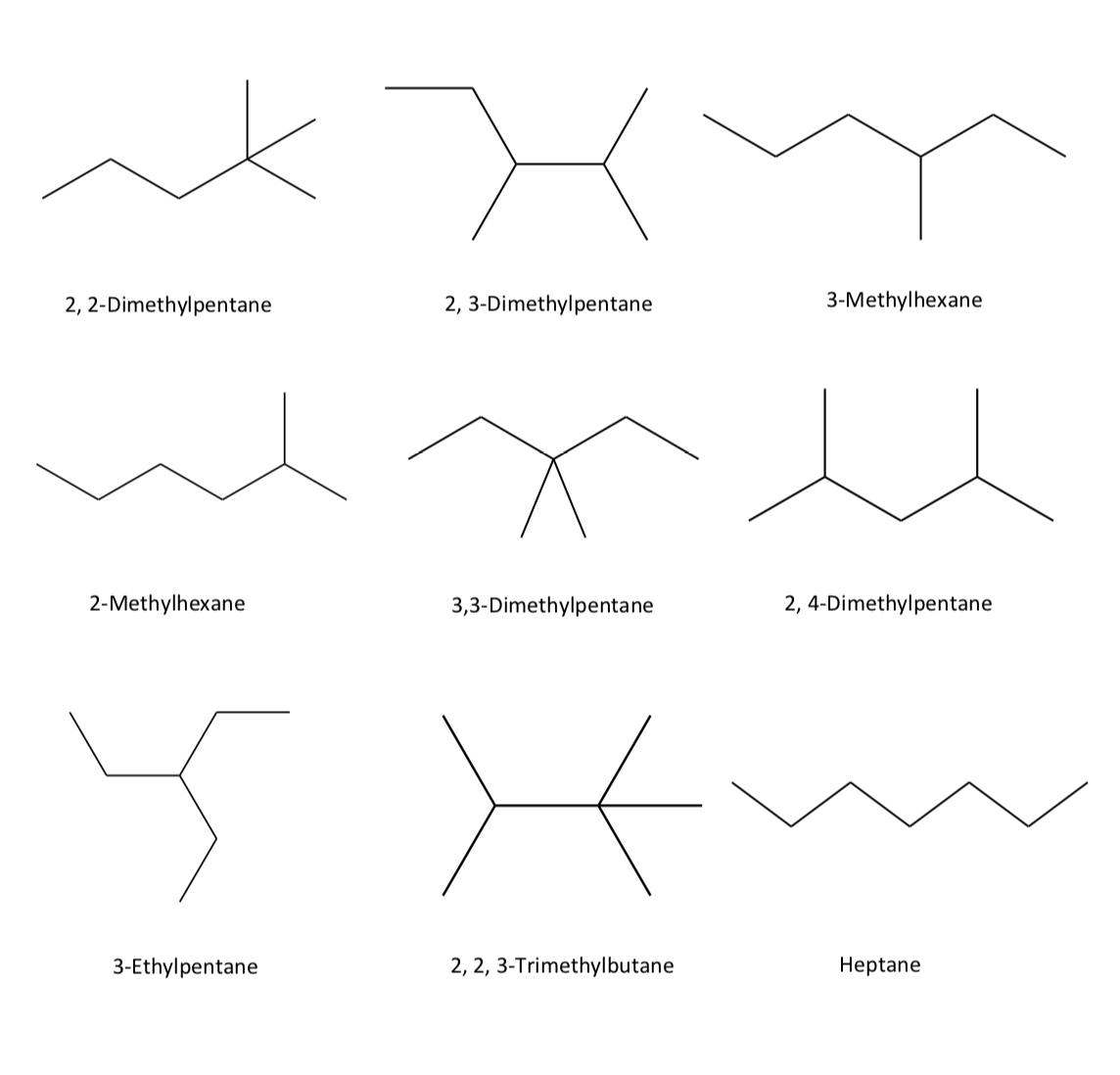}
    \caption{{\bf Heptane Isomers}
      Created graphs and corresponding isomers for Heptane ($C_{7}H_{16}$) Left: Returned graphs with degree sequences. Right: Isomers of Heptane.}
    \label{heptane_graphs}
  \end{figure}

  Using Python packages \texttt{NumPy}~\cite{numpy}, \texttt{D-Wave Ocean}~\cite{ocean}, \texttt{Sympy}~\cite{sympy}, \texttt{NetworkX}~\cite{networkx}, and \texttt{Matplotlib}~\cite{matplotlib} and the D-Wave 2000Q hardware, all structural isomers for Butane ($C_{4}H_{10}$), Pentane ($C_{5}H_{12}$), Hexane ($C_{6}H_{14}$), Heptane ($C_{7}H_{16}$), Octane ($C_{8}H_{18}$), and Nonane ($C_{9}H_{20}$) were identified. These molecules have 2, 3, 5, 9, 18, and 35 isomers, respectively. Fig.~\ref{heptane_graphs} shows the returned graphs and their corresponding isomers for Heptane ($C_{7}H_{16}$).
  
  \par
  
  Without using QUBO perturbation and reverse annealing, it was found that 10,000 samples were sufficient to find all isomers for Butane ($C_{4}H_{10}$) and Pentane ($C_{5}H_{12}$), but the larger alkanes often needed well over 50,000 samples in order to be fully captured. Information evaluating and describing the results is given below.
  
  \par
  Fig.~\ref{hamming_distances} gives information on the Hamming distances of all of these isomers. The Hamming distance between two isomers is the number of bit flips that must be made in order to turn one isomer into the other. The left figure shows all pairwise Hamming distances for a given $n$, and the right figure shows the minimum Hamming distance to each isomer for a given $n$. As can be seen, while the pairwise Hamming distances tend to follow a fairly wide distribution, almost every isomer has another isomer within the minimum possible Hamming distance (4).
  
  \begin{figure}[!h]
    \centering
    \includegraphics[width=0.48\linewidth]{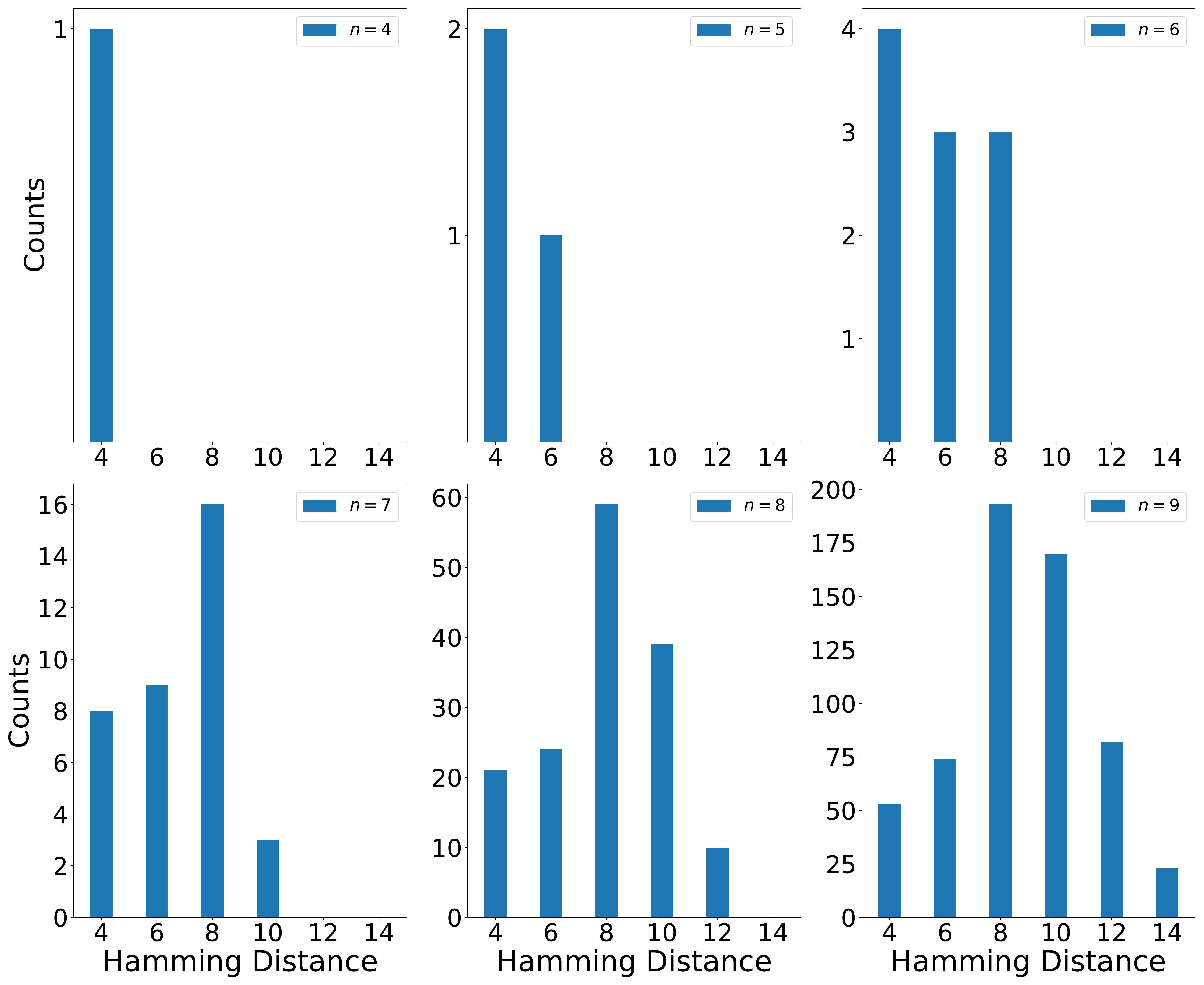}
    \includegraphics[width=0.48\linewidth]{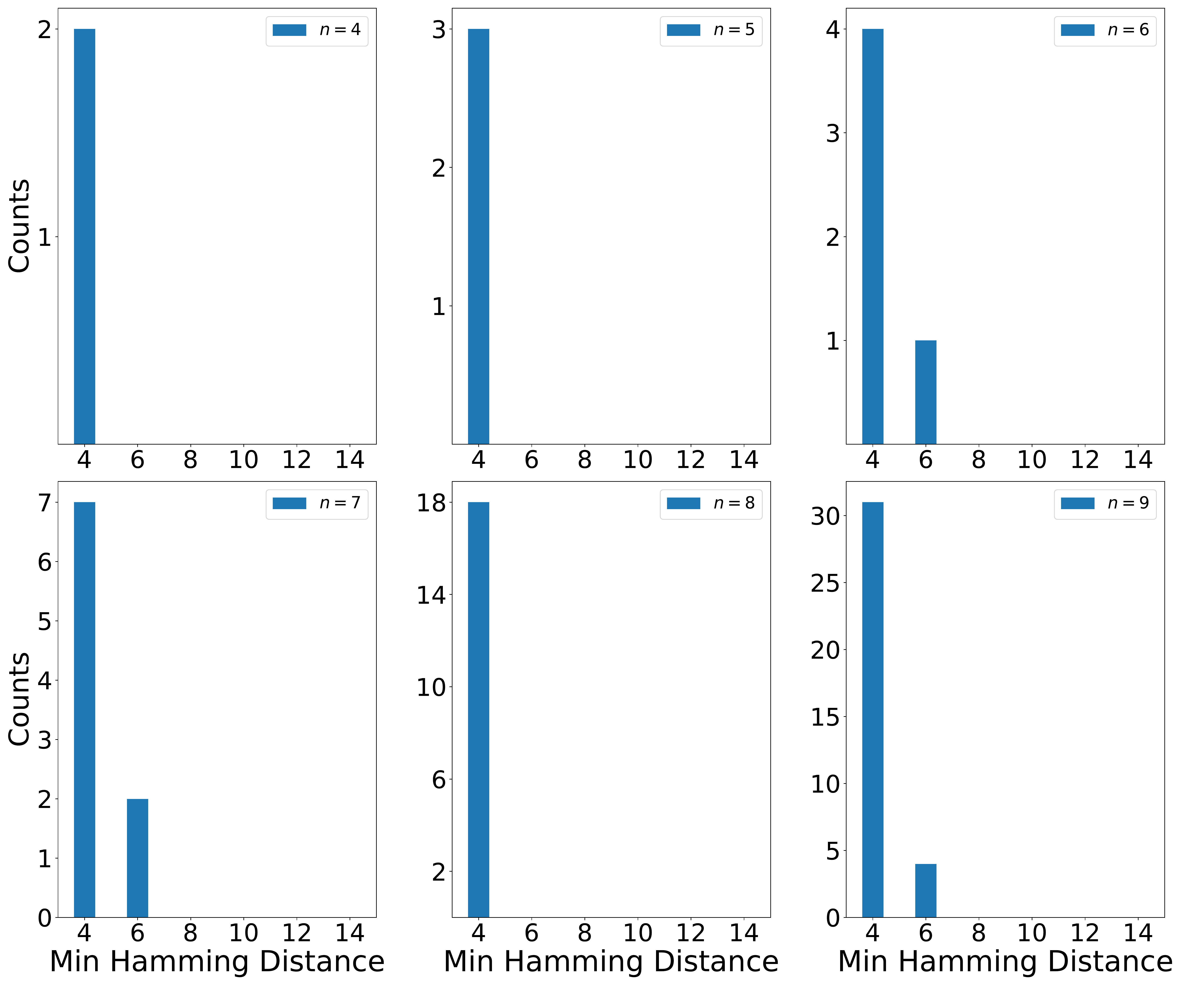}
    \caption{{\bf Measures of Hamming Distances.}
      Left: All pairwise Hamming distances, Right: Minimum Hamming distance to each isomer}
    \label{hamming_distances}
  \end{figure}
  
  \par
\begin{figure}[!h]
    \centering
    \includegraphics[width=0.495\linewidth]{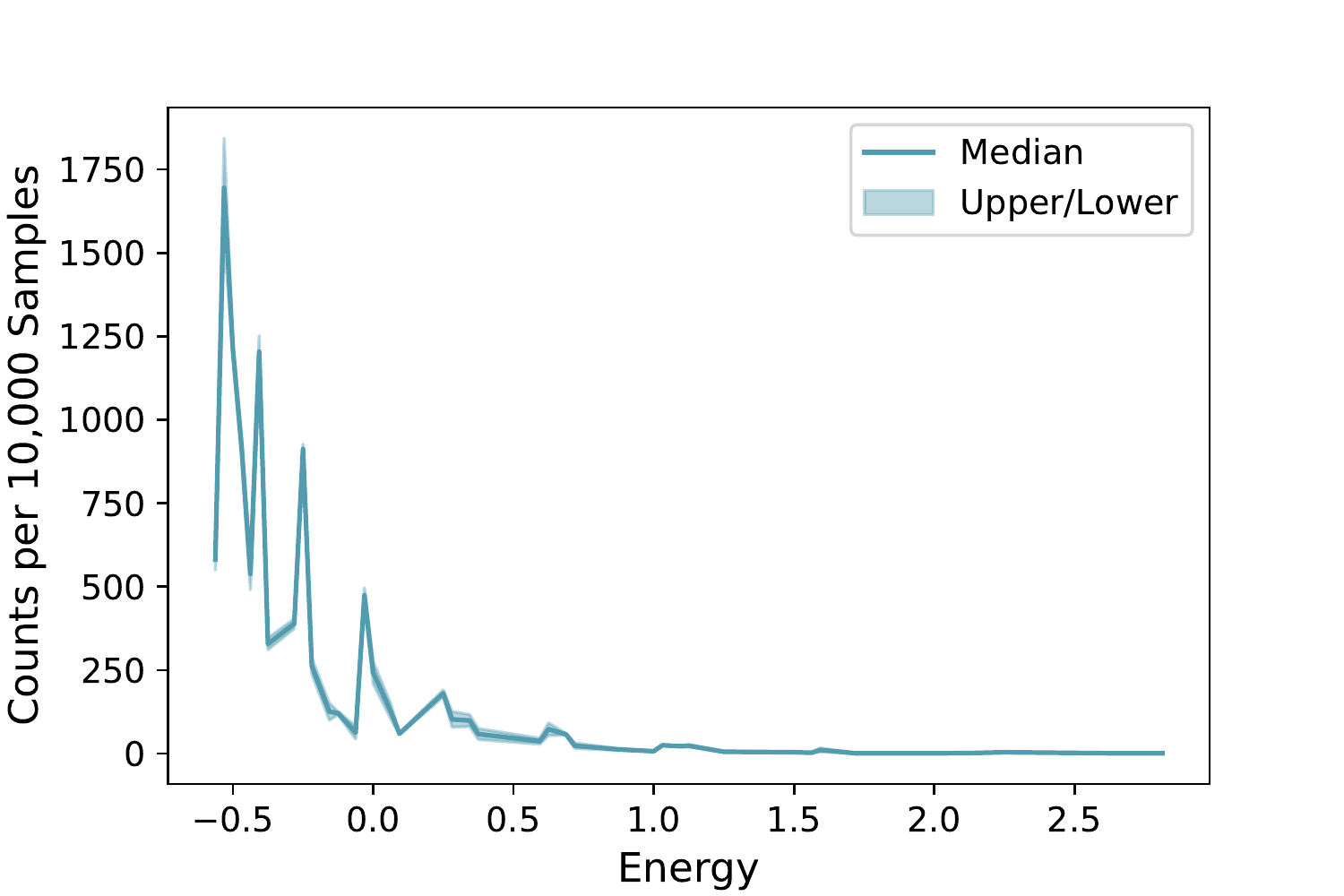}
    \includegraphics[width=0.495\linewidth]{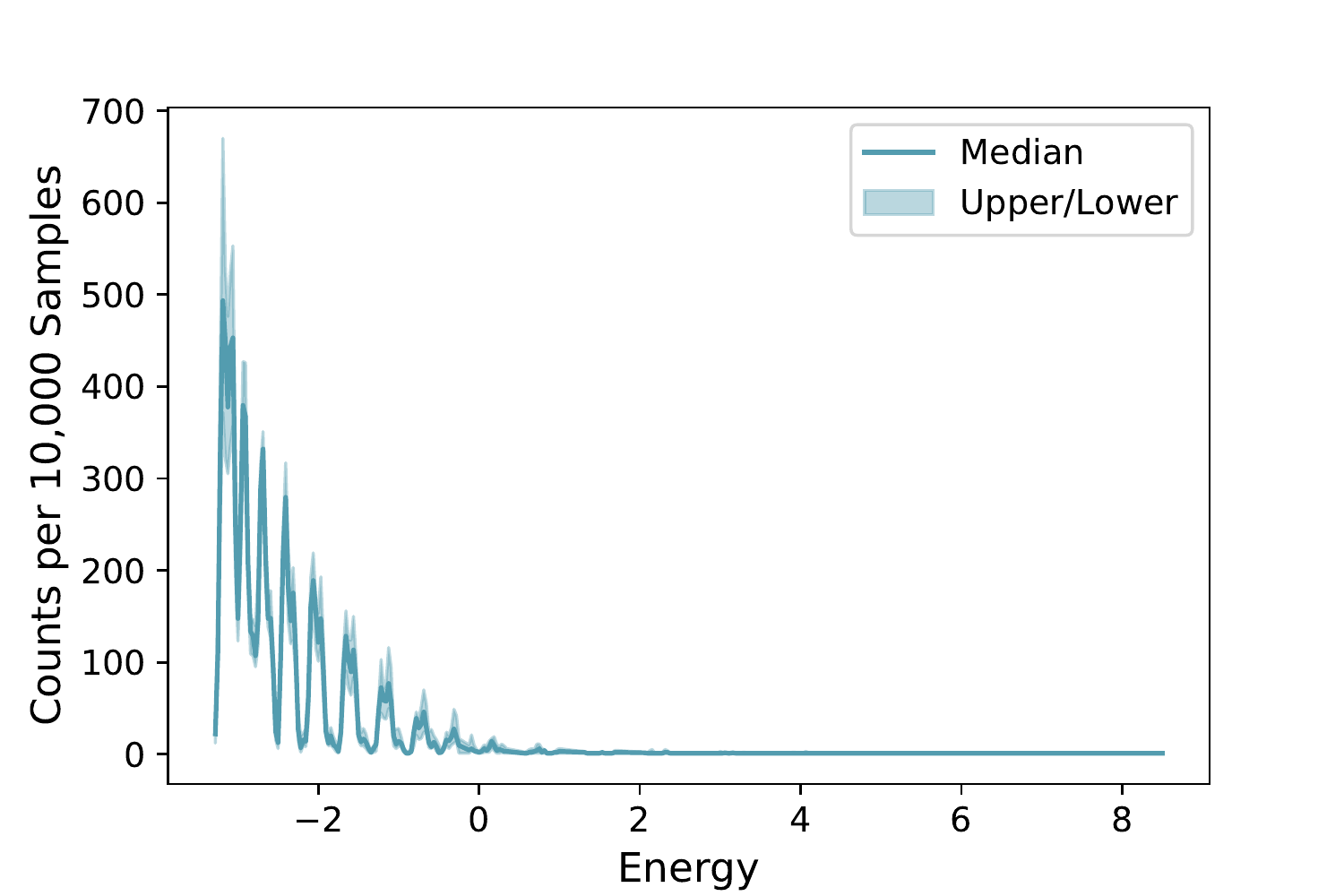}
    \caption{{\bf Number of Results Returned for Each Energy.}
      Number of results out of 10,000 samples returned at each energy. Left: Butane ($C_{4}H_{10}$), Right: Heptane ($C_{7}H_{16}$)}
    \label{num_p_energy}
  \end{figure}
\begin{figure}[!h]
    \centering
    \includegraphics[width=0.495\linewidth]{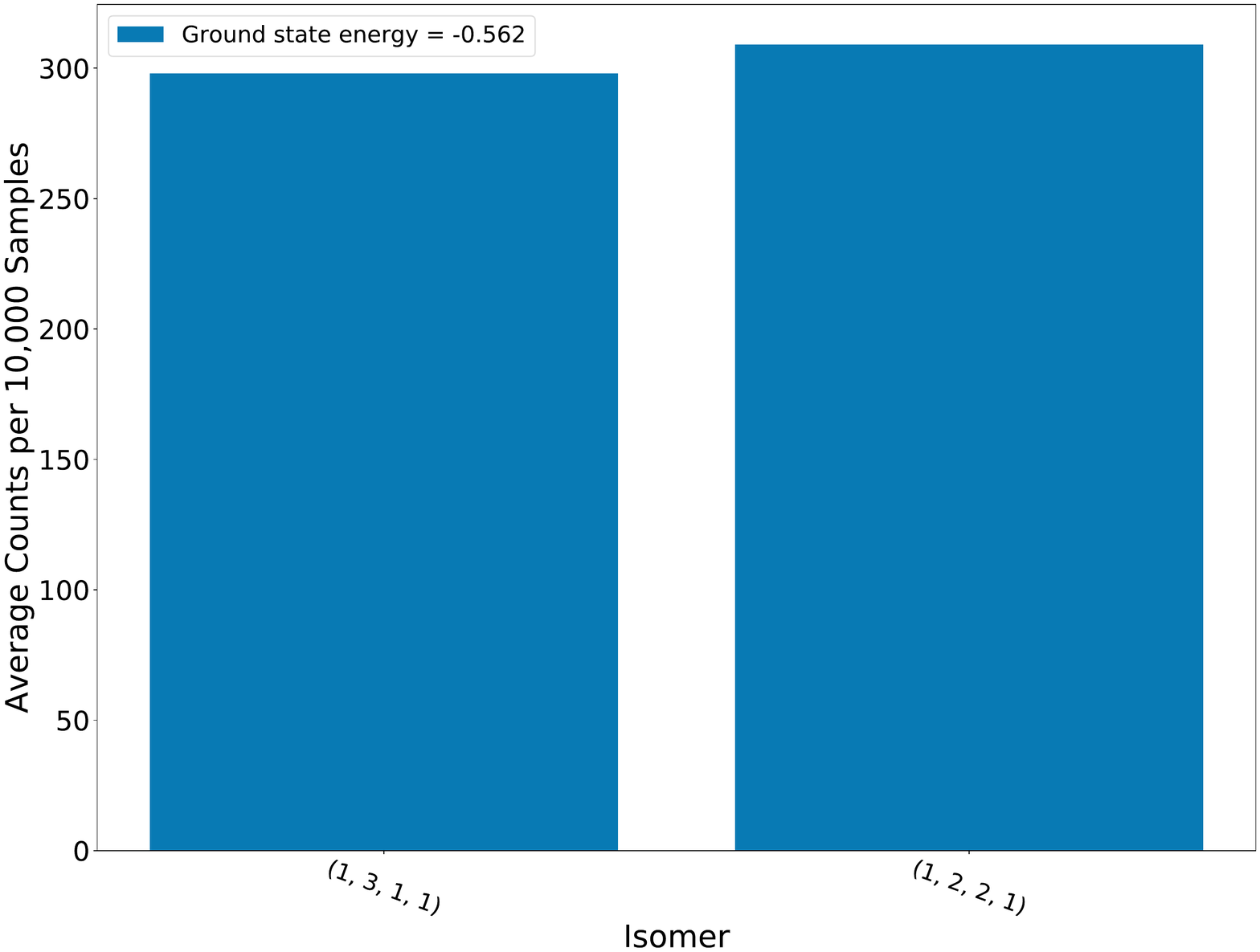}
    \includegraphics[width=0.495\linewidth]{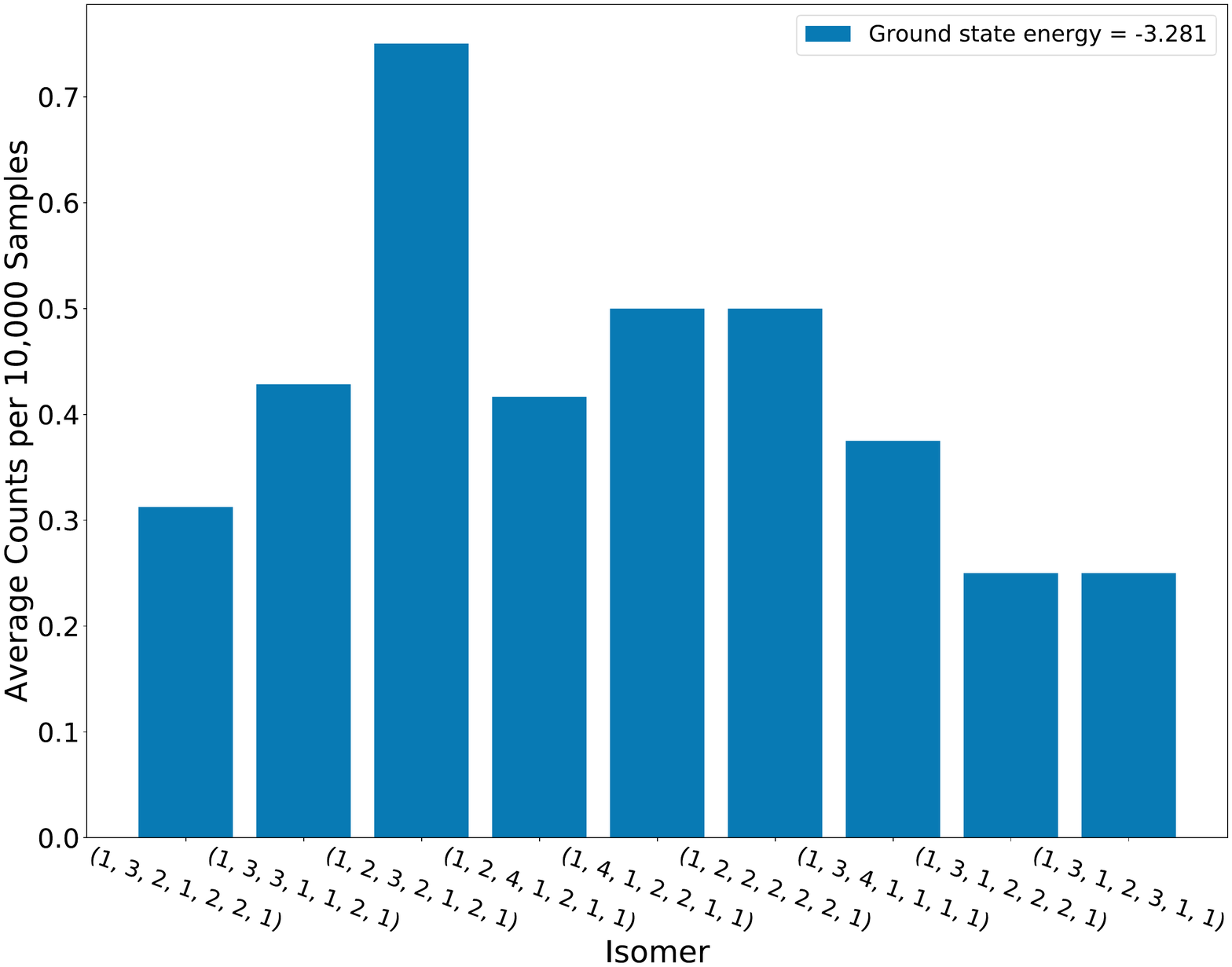}
    \caption{{\bf Distribution of Returned Isomers.}
      Average number of times each isomer was returned per 10,000 samples. Left: Butane ($C_{4}H_{10}$), Right: Heptane ($C_{7}H_{16}$)}
    \label{avg_count}
  \end{figure}

  Fig.~\ref{num_p_energy} and \ref{avg_count} give information on the frequency with which isomers are found for Butane ($C_{4}H_{10}$) and Heptane ($C_{7}H_{16}$). In both figures, it is easily seen that isomers of Butane, with only 4 carbon atoms, are much more easily found than those of Heptane ($C_{7}H_{16}$). Fig.~\ref{num_p_energy} (left) demonstrates that ground state results, i.e. those that violate no constraints, are found several hundred times per 10,000 anneals when Butane ($C_{4}H_{10}$) is being investigated. However, as the Fig.~\ref{num_p_energy} (right) shows, only a handful are sampled for Heptane ($C_{7}H_{16}$). Fig.~\ref{avg_count} (left) shows that each isomer is, on average, found several hundred times per 10,000 samples whereas the right plot shows that isomers of Heptane ($C_{7}H_{16}$) are generally found less than once per 10,000 samples.
  
  \par
\begin{figure}[!h]
    \centering
    \includegraphics[width=0.495\linewidth]{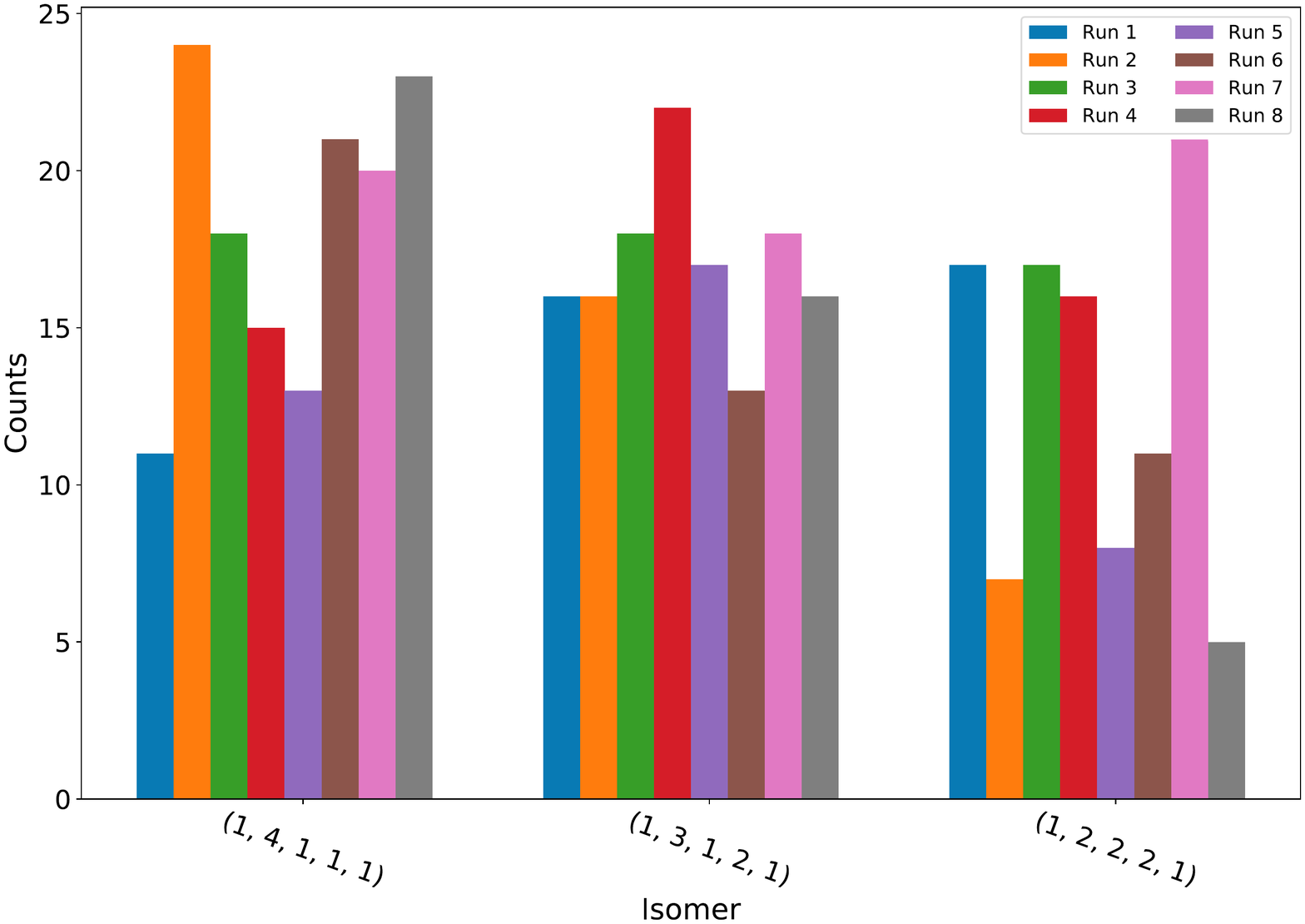}
    \includegraphics[width=0.495\linewidth]{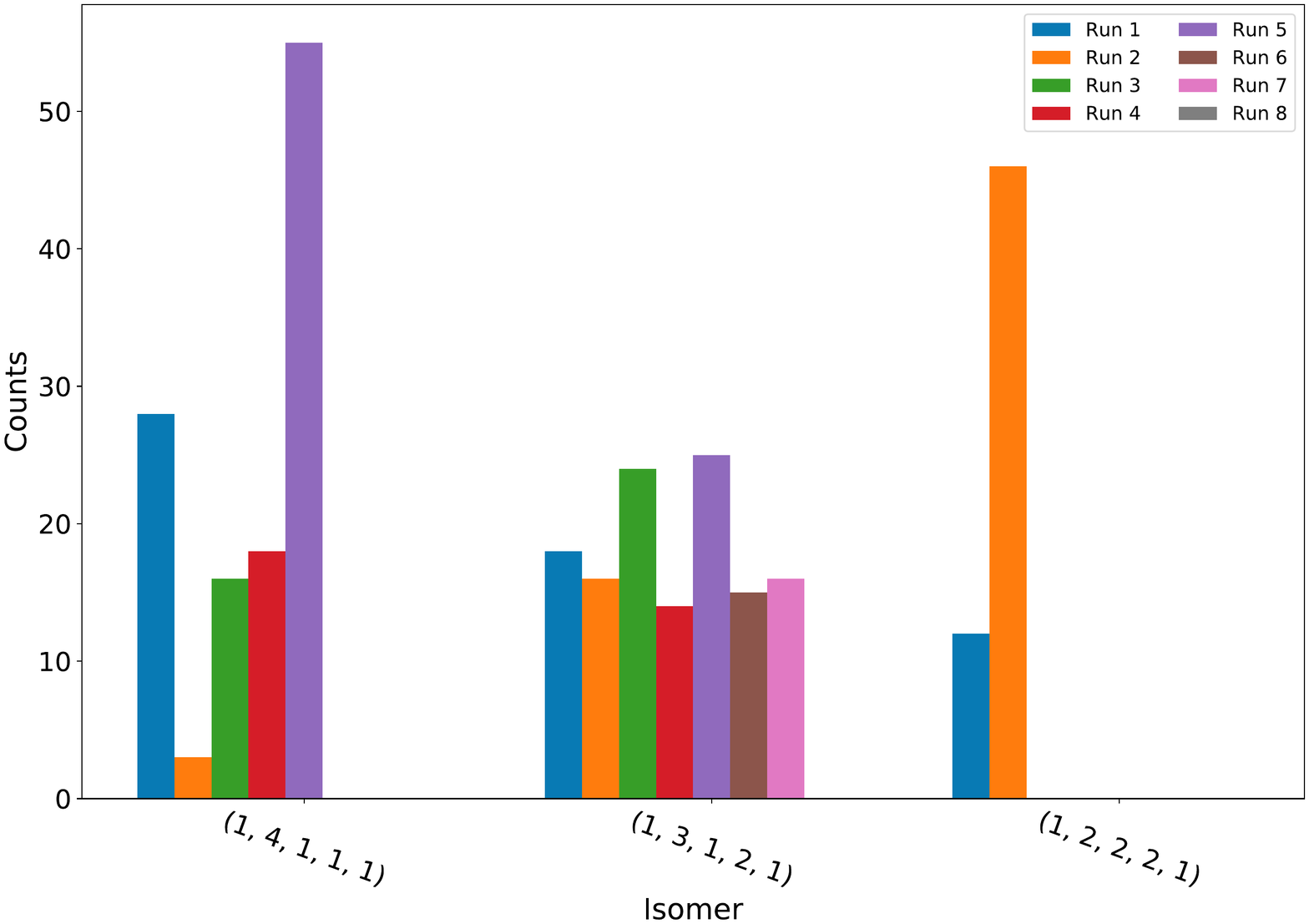}
    \caption{{\bf Sequential Distributions of Results.}
      Distributions of returned isomers of Pentane ($C_{5}H_{12}$) after each run of 10,000 samples. Left: Not using QUBO perturbation, Right: Using QUBO perturbation.}
    \label{perturbation}
  \end{figure}

  Our sample reduction methods were also explored and evaluated. Perturbing the QUBO clearly had an effect on the distribution of the returned results. Fig.~\ref{perturbation} gives the distributions of the returned isomers with (left) and without (right) perturbing the QUBO for Pentane ($C_{5}H_{12}$) after 10,000 samples using $\lambda=5(10^{-5})$. The distributions for the non-perturbed QUBO runs, Fig.~\ref{perturbation} (left), are somewhat uniform. Every isomer is found during each iteration, and the isomers are roughly returned at the same rate. The randomness of the annealing will always introduce some fluctuations. However, these fluctuations are not too large and tend to settle back to normal by the next iteration. This is starkly contrasted by Fig.~\ref{perturbation} (right). This shows the distributions when QUBO perturbation is used and as can easily be seen, this method drastically changes the results for subsequent runs. The isomer that is sampled the most frequently for a given iteration is typically sampled significantly fewer times during the next iteration. Eventually, an isomer that has been sampled the most frequently is penalized to the extent that it is never sampled again. By the final iteration, each isomer has at some point been the most frequently sampled. This seems to drive the QUBO so far away from them that none of them are sampled.

  This QUBO perturbation technique and reverse annealing were tried alone and in tandem. It was found that, by themselves, each typically led to a reduction in the number of samples needed, but combining them decreased the number of samples even more significantly. The effect of these methods on the search for isomers of Heptane ($C_{7}H_{16}$) was measured by finding the number of iterations of 10,000 samples that were necessary to find all isomers. As this is not a constant number, the experiment was repeated 25 times for each of the four methods (only forward annealing, forward annealing with QUBO perturbation, reverse annealing, and reverse annealing with QUBO perturbation). The results are shown in Table~\ref{expected_runs}. This gives the average and median number of runs needed to find all isomers for each different technique over the 25 runs. As it shows, both QUBO perturbation and reverse annealing separately outperform a typical forward anneal, but combining the two gives the largest reduction in the number of samples needed.

  \begin{table}[!ht]
    \centering
    \caption{
      {\bf Number of Runs to Find All Heptane Isomers}}
    \begin{tabular}{|c|c|c|c|c|}
      \hline
      & \multicolumn{1}{|l|}{\bf FA} & \multicolumn{1}{|l|}{\bf FA + QP} & \multicolumn{1}{|l|}{\bf RA} & \multicolumn{1}{|l|}{\bf RA + QP} \vline \\ \thickhline
      Mean & 9.68 & 8.44 & 8.04 &  6.56\\ \hline
      Median & 9 & 7 & 8 &  6 \\ \hline
    \end{tabular}
    \begin{flushleft} 
      Average and median number of samples (in 10,000s) needed to find all isomers of Heptane ($C_{7}H_{16}$) using forward annealing (FA), forward annealing and QUBO perturbation (FA + QP), reverse annealing (RA), and reverse annealing and QUBO perturbation (RA + QP). $s^{*} = 0.5$, $h = 85\mu s$, $\lambda=5(10^{-6})$.
    \end{flushleft}
    \label{expected_runs}
  \end{table}

  \par
\begin{figure}[!h]
    \centering
    \includegraphics[width=0.8\linewidth]{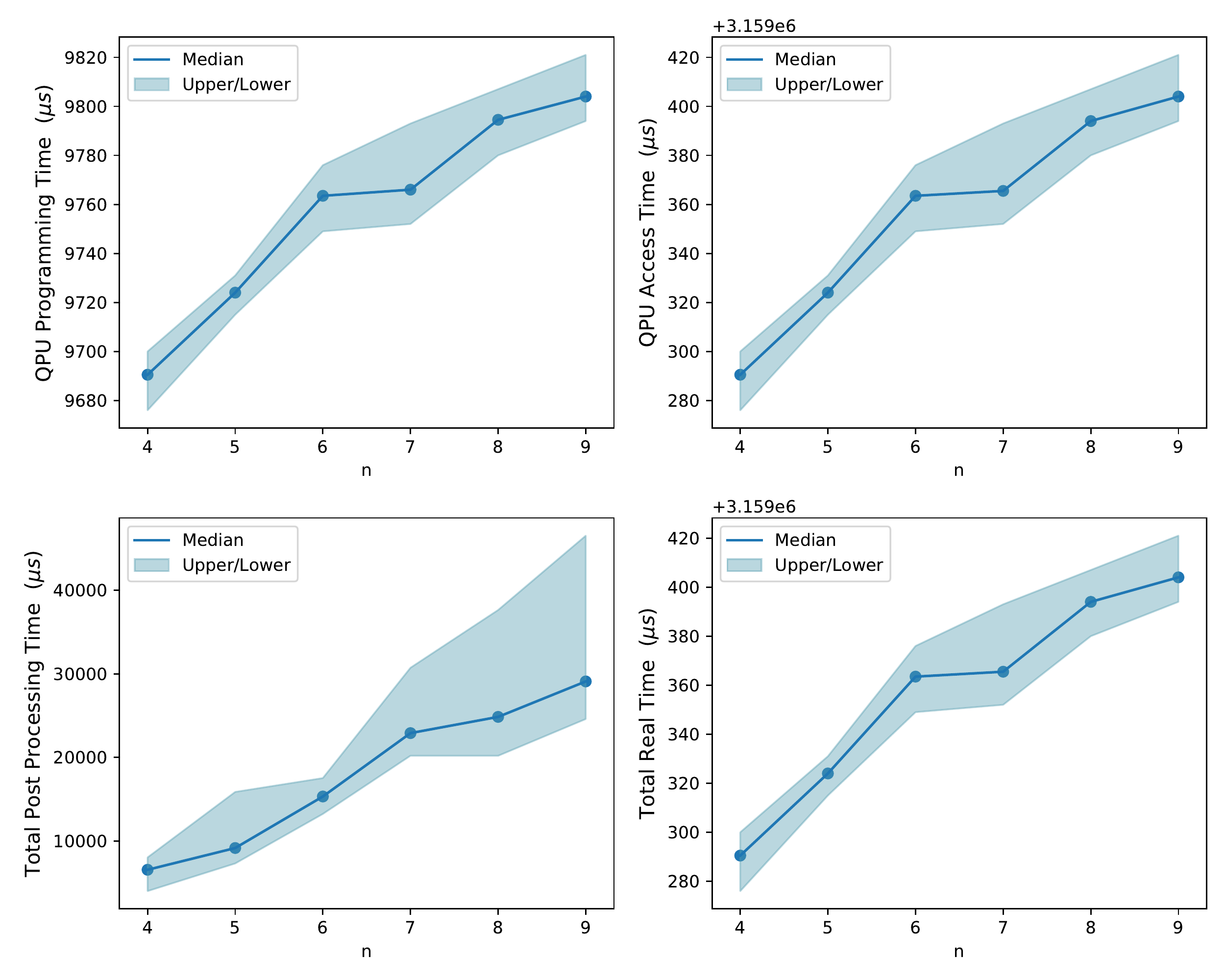}
    \caption{{\bf Benchmark Times.}
      Time taken per 10,000 samples for $4\leq n \leq 9$. Upper Left: QPU programming time, Upper Right: QPU access time, Lower Left: Total post processing time, Lower Right: Total real time. Notice the scale difference on the right two panels.}
    \label{times}
  \end{figure}

  When the runtime scaling was evaluated, it was found that the time taken to generate 10,000 samples grows fairly linearly with $n$, with each additional carbon atom adding roughly 20 microseconds to the total runtime, seen in Fig.~\ref{times}. This happens despite the fact that the number of parameters grows quadratically. QPU access time is shown in Fig.~\ref{times} (upper right) includes everything that is done on the QPU: QPU programming time and QPU sampling time. QPU programming time, which is shown in Fig.~\ref{times} (upper left), measures how long it takes to initialize the problem on the QPU, a procedure that is only done once per 10,000 samples. QPU sampling time includes the time it takes to perform and readout all anneals, with delays in between subsequent samples to allow for the system to return to its initial temperature~\cite{dwave_time}. After all samples are collected, they then undergo post-processing in an attempt to improve the quality of the solutions~\cite{postprocessing1, postprocessing2}. Total post-processing time is shown in Fig.~\ref{times} (lower left). All of this put together is the total real time and is shown in Fig.~\ref{times} (lower right). As can be seen, QPU access is by far the dominant time consumer of the experiments. 
  
  \par
  Finally, there was also some attempt to use IBM Q's \texttt{Qiskit} to implement the search on their hardware~\cite{qiskit}. Current attempts on the IBM Q simulator using this method have been able to identify all isomers of Butane ($C_{4}H_{10}$) and calculate the correct ground state energy of the QUBO Hamiltonian.  However, 2-methylpropane (an isomer of Butane $C_{4}H_{10}$) is found very often, typically well over five hundred counts per 8,192 samples. This generally makes it one of the three most common results. However, unbranched Butane ($C_{4}H_{10}$) occurs much more rarely.
  
  \section*{Discussion}
  
  These results are a proof of concept that quantum isomer search using a QUBO formulation is a valid method. With this approach all isomers for all alkanes with fewer carbon atoms than Decane ($C_{10}H_{22}$) were identified. However, as the number of carbon atoms grows, it becomes more and more essential to take more samples. As shown in Table~\ref{expected_runs}, combining reverse annealing with our method of perturbing the QUBO after every iteration of 10,000 samples drastically decreased the number of samples required to find all isomers. Therefore, we can speculate that perturbing the QUBO and reverse annealing are important methods that may significantly help expand the search space, decrease the runtime, and facilitate the complete identification of isomers for larger molecules. 
  
  \par
  It is important to note that the necessity of increasing the number of samples as the problem size grows is indicative of imperfect hardware. This scaling is not an inherent part of the quantum isomer search algorithm. Increasing the problem size decreases the probability of finding a successful answer due to annealing error and imperfect hardware~\cite{error}. It is not surprising that more runs are needed for larger molecules, especially when the quadratic scaling of parameters and the limited connectivity of the D-Wave 2000Q's chimera graph are taken into account. This is a large contributor to the need for more sampling for larger molecules and comes from imperfect hardware rather than the scaling of the algorithm itself.
  
  \par
\begin{figure}[!h]
    \centering
    \includegraphics[width=0.495\linewidth]{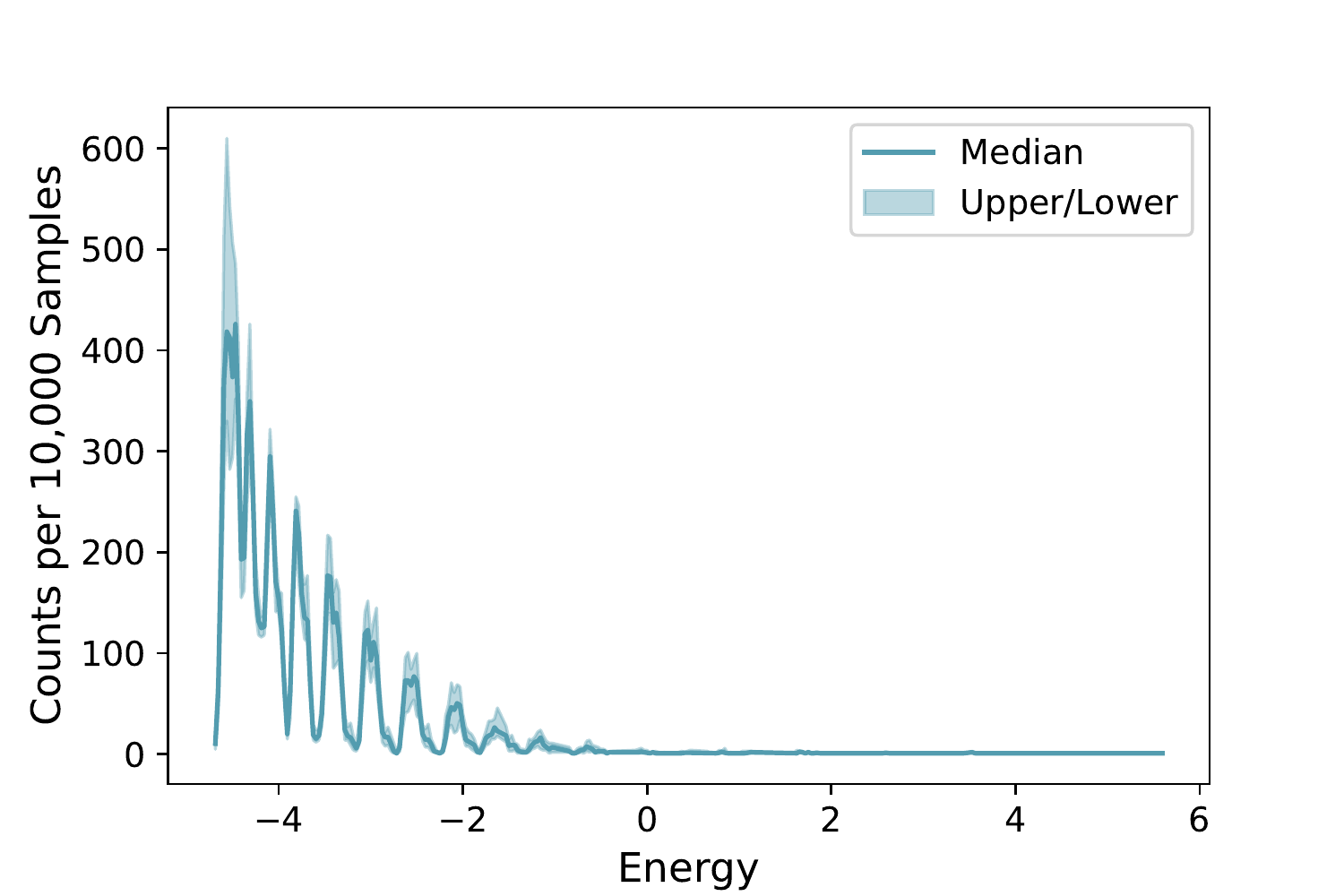}
    \includegraphics[width=0.495\linewidth]{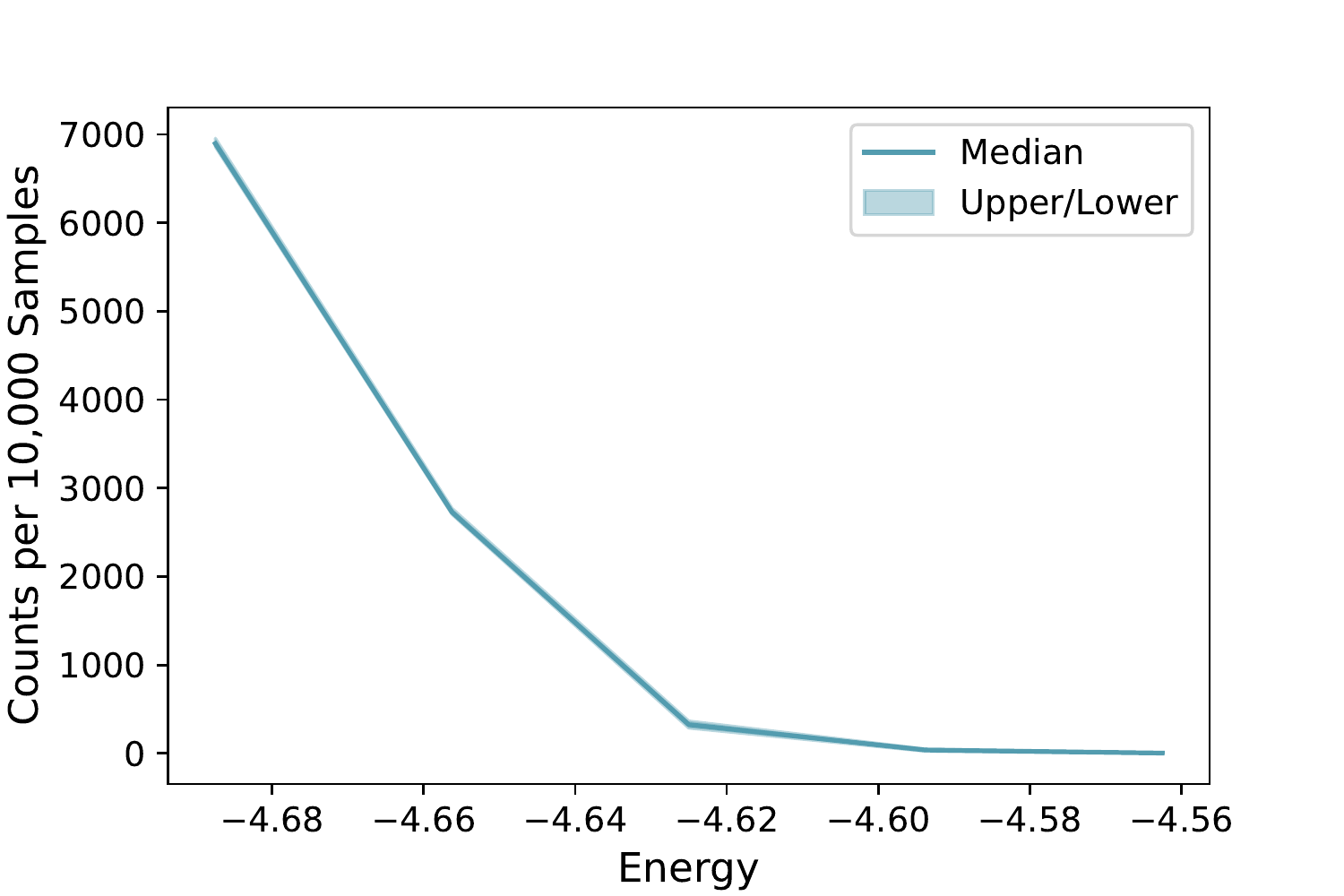}
    \caption{{\bf Number of Results Returned for Each Energy}
      Left: using quantum annealing, Right: using simulated annealing}
    \label{simulated_samples}
  \end{figure}

  Evidence for this can be seen in Fig.~\ref{simulated_samples}. This figure compares the number of samples returned for each energy when the isomers of Octane ($C_{8}H_{18}$) are searched for. As is easily seen, the number of ground state samples is significantly lower for the quantum annealer (left) when compared to the simulated annealer (right). This indicates that the algorithm is working correctly, but the hardware is limiting its performance. Furthermore, every isomer was able to be found within 10,000 samples when simulated annealing was used, regardless of the size of the molecule. This also indicates the fact that the imperfect hardware is limiting the performance and is responsible for the increase in the number of samples needed for larger molecules to some extent. 
  
  \par
  We speculate, however, that as the hardware's performance increases, it may be the case that the quantum annealer will outperform the simulated annealer, particularly in runtime. This is due to the linear scaling of the quantum runtime with problem size, whereas simulated annealing has been found to scale exponentially in some cases~\cite{reverse_anneal}. In fact, some problems with 945 variables were found to run over $10^{8}$ times slower when simulated annealing was used~\cite{qa_scaling}. Embedding 13 carbons, i.e. searching for isomers of Pentadecane ($C_{15}H_{32}$), requires 946 variables. Because this molecule has fewer than 18 carbons, it is small enough to already be embedded on D-Wave 2000Q. Therefore, this enormous potential decrease in runtime may occur for isomers that are small enough to be embedded on current hardware. However, as discussed earlier, even though a molecule with this size can be embedded, the hardware's error makes searching for the 4,347 isomers impractical. As a result, this runtime decrease will have to wait for better hardware. We make no claim that the potential decrease in runtime will be of the same magnitude as that found in~\cite{qa_scaling}. However, the complexity of the isomer search problem along with the sheer magnitude of the runtime decrease that was demonstrated suggest that some decrease in runtime should be expected when quantum isomer search is applied to larger molecules.

  \par
  Degeneracy is an additional issue that pertains to the need for thorough sampling. As described earlier, the global minimum is degenerate in that there are multiple solutions that satisfy all constraints and all of these need to be sampled in order to ensure that all isomers are found. This necessitates a very thorough exploration of the search space. Such a requirement is the reason that QUBO perturbation was added. The results seem to suggest that this would encourage a wider exploration of the search space. The complication is furthered by the fact that for several isomers, there are multiple permutations of a degree sequence that lead to identical graphs. Therefore, in a way the degeneracy is two-fold. There are multiple valid and distinct global minima, but there are also multiple valid yet identical global minima. The degeneracy will only increase as the problem size increases and is perhaps one of the largest limitations.
  
  \par
  The degeneracy is further complicated by the Hamming distances between the isomers. The Hamming distance is a measure of how many modifications need to be made to a result in order to transform it into another result. Because of the one hot encoding of the degrees, changing the degree of one carbon requires 2 bit flips. The constraint pertaining to the sum of the degrees means that if one degree is changed, at least one other degree must be changed and at most $(n-2)$ degrees. The total number of carbons embedded may be changed. Therefore, the Hamming distance between any two isomers of a given $n$ is strictly within 4 and $2(n-2)$, inclusive. As can be seen in the left panel of Fig.~\ref{hamming_distances}, the pairwise distribution of Hamming distances follows this pattern. However, when the minimum Hamming distance to a given isomer is calculated, as is shown in the right panel of Fig.~\ref{hamming_distances}, it is seen that for almost every isomer for all $n$ another isomer can be made by changing only 2 degrees. Therefore, while any two isomers may be far apart, almost every isomer has another isomer quite close by. 
  
  \par
  When measured in Hamming distances, the isomers can form very close clumps of ground state results. It is possible that this property helps simulated annealing deal with degeneracy in an effective way. Once a given clump is found, the other isomers can be found by only flipping a handful of bits. The implications for quantum annealing are less clear. It is entirely possible that a clump may be found, but not all isomers within that clump are found because the quantum annealing explores so vastly that it may quickly leave the clump. Alternatively, it is possible that the annealer would be drawn to larger clumps, i.e. isomers from the clumps that contain many molecules with small Hamming distances would be more likely to be visited and explored. It is possible that this is an issue that can be addressed to some extent by the introduction of reverse annealing. Its ability to do local searches surrounding the candidate solution given by forward annealing may allow it to explore a given clump more fully. Furthermore, QUBO perturbation may help the annealing explore between clumps by driving the search away from clumps with answers that were already visited. This complementary combination of exploring within and between clumps introduced by reverse annealing and QUBO perturbation may be the reason that combining the two methods is so effective in terms of decreasing the number of samples needed to find all isomers. There very well may be other implications that we have not brought up, so this may be an interesting direction for further research on degenerate problems.
  
  \par
  On D-Wave 2000Q, the sampling of a QUBO that grows quadratically in the number of parameters can be done in linear time. Despite the fact that more samples are required for larger molecules, the linear scaling of sampling time is an important quality. When combined with the significant reduction in samples needed due to QUBO perturbation and reverse annealing it becomes an encouraging sign for the feasibility of applying this method to larger molecules.
  
  \par
  Even though the current results on the IBM Q hardware are not competitive with those from D-Wave 2000Q, the rapidly increasing performance and growing number of qubits on this and other gate-based machines makes this direction a promising avenue for further research. However, variational techniques on gate-based machines may not have the linear runtime scaling that is found when quantum annealing is used. This is because these techniques will likely use QUBO Hamiltonians that come with a quadratically scaling number of terms due to the quadratic scaling of the problem. Because the expectation value of all of these terms must be taken when a technique such as VQE is used, this will likely result in a runtime that scales closer to quadratic than linear. Furthermore, initial results do not seem to indicate that these variational techniques can handle degeneracy as effectively as quantum annealing. However, it seems that this limitation may be addressed to some extent by QUBO perturbation as well as other methods including low-energy subspace sampling using something akin to a subspace-search variational quantum eigensolver~\cite{SSVQE}.
  
  \section*{Conclusion}
  
  We have demonstrated that quantum isomer search using the QUBO formulation is possible and effective. With our approach, the sampling time 
  grows linearly with the number of carbon atoms. All isomers for all alkanes with fewer carbon atoms than Decane ($C_{10}H_{22}$) were identified and enumerated using this approach on the D-Wave 2000Q system. 
  Alkanes with fewer carbon atoms than Nonadecane ($C_{19}H_{40}$) can be embedded directly into the D-Wave 2000Q for the quantum isomer search. 
  However, the next-generation D-Wave with 5000 qubits is coming soon. Along with its decrease in noise and significant increase in the number of physical qubits, it will also feature a more connected Pegasus graph in which each physical qubit is connected to 15 others rather than only 6~\cite{pegasus}. This combination will allow the isomers of much larger molecules to be searched for. It is likely that as the problem size increases, the importance of the significant sample reduction and wider exploration of the search space made possible by perturbing the QUBO and adding reverse annealing will quickly grow.
  \par
  The natural next step of this problem is to implement it on a gate-based quantum computer. Variational methods on these computers can also solve the Ising problem, so quantum isomer search is possible on those machines. However, all available gate-based quantum computers have significantly fewer qubits than D-Wave 2000Q, so they can only search for isomers of relatively small molecules. Regardless, the number of qubits that these machines have is quickly growing as their noise is decreasing, so it is a promising direction of future work.

  \section*{Acknowledgments}
  The authors would like to acknowledge the NNSA's Advanced Simulation and 
  Computing (ASC) program at Los Alamos National Laboratory
  (LANL) for use of their Ising D-Wave 2000Q quantum computing resource. 
  LANL is operated by Triad National Security, LLC, for the National
  Nuclear Security Administration of U.S. Department of Energy (Contract No. 89233218NCA000001).
  The authors would also like to acknowledge D-Wave Systems Inc. for the use of their lower-noise D-Wave 2000Q from D-Wave Leap.
  Jason Terry was funded by the LANL Quantum Computing Summer School (QCSS) 2019. The QCSS is
  sponsored by the LANL Information Science and Technology Institute (ISTI). 
  This research was also supported in part by an appointment by the National Science Foundation (NSF) 
  Mathematical Sciences Graduate Internship (MSGI) Program sponsored by the NSF Division of Mathematical Sciences. Prosper Akrobotu's contribution to this work was funded by NSF-MSGI. This program is administered by the Oak Ridge Institute for Science and Education (ORISE) through an interagency agreement between the U.S. Department of Energy (DOE) and NSF. ORISE is managed for DOE by ORAU. All opinions expressed in this paper are the author's and do not necessarily reflect the policies and views of NSF, ORAU/ORISE, or DOE.
  Christian Negre's and Sue Mniszewski's contributions to this research have been funded by the
  LANL ISTI and Laboratory Directed Research and Development (LDRD). 
  Assigned: Los Alamos Unclassified Report 19-26724.
  
  \nolinenumbers
  
  \bibliographystyle{plainnat}
  \bibliography{paper}
  
\end{document}